\documentclass[pdflatex,sn-basic,iicol]{sn-jnl}

\jyear{2023}%

\theoremstyle{thmstyleone}%
\theoremstyle{thmstyletwo}%

\theoremstyle{thmstylethree}%

\newcommand\fs{\mbox{$.\!\!^{\mathrm s}$}}%
\newcommand\fdg{\mbox{$.\!\!^\circ$}}%

\newcommand\farcs{\mbox{$.\!\!^{\prime\prime}$}}%
\let\farcs\farcs 
\newcommand\micron{\mbox{$\mu$m}}%

\newcommand{\aj}{Astron. J.}   
\newcommand{\apj}{Astrophys. J.}   
\newcommand{\apjl}{Astrophys. J. Lett.}   
\newcommand{\apjs}{Astrophys. J. Suppl. Ser.}   
\newcommand{\aap}{Astron. Astrophys.}   
\newcommand{\aaps}{Astron. Astrophys. Suppl.}   
\newcommand{\mnras}{Mon. Not. R. Astron. Soc.}   
\newcommand{\nat}{Nature} 
\newcommand{\pnas}{Proc. Natl Acad. Sci. USA}   
\newcommand{\pasp}{Publ. Astron. Soc. Pac.}   
\newcommand{\solphys}{Sol. Phys.}   

\newcommand*{\figuretitle}[1]{%
    {\centering
    \textbf{#1}
    \par\medskip}
}

\begin{document}

\title[]{
Spatially resolved imaging of the inner Fomalhaut disk using JWST/MIRI}

\subtitle{}


\author*[1]{\fnm{Andr\'as} \sur{G\'asp\'ar}}\email{agaspar@arizona.edu}

\author[1]{\fnm{Schuyler Grace} \sur{Wolff}}

\author[1]{\fnm{George H.} \sur{Rieke}}

\author[1]{\fnm{Jarron M.} \sur{Leisenring}}

\author[1]{\fnm{Jane} \sur{Morrison}}

\author[1]{\fnm{Kate Y. L.} \sur{Su}}

\author[2]{\fnm{Kimberly} \sur{Ward-Duong}}

\author[3]{\fnm{Jonathan} \sur{Aguilar}}

\author[4]{\fnm{Marie} \sur{Ygouf}}

\author[4]{\fnm{Charles} \sur{Beichman}}

\author[5]{\fnm{Jorge} \sur{Llop-Sayson}}

\author[4]{\fnm{Geoffrey} \sur{Bryden}}

\affil*[1]{\orgdiv{Steward Observatory and the Department of Astronomy}, \orgname{The University of Arizona}, \orgaddress{\street{933 N Cherry Ave}, \city{Tucson}, \postcode{85719}, \state{AZ}, \country{USA}}}

\affil[2]{\orgdiv{Department of Astronomy}, \orgname{Smith College}, \orgaddress{\street{2 Tyler Ct}, \city{Northampton}, \postcode{01063}, \state{MA}, \country{USA}}}

\affil[3]{\orgdiv{Space Telescope Science Institute}, \orgname{AURA}, \orgaddress{\street{3700 San Martin Drive}, \city{Baltimore}, \postcode{21218}, \state{MD}, \country{USA}}}

\affil[4]{\orgdiv{Jet Propulsion Laboratory}, \orgname{California Institute of Technology}, 
\orgaddress{\street{4800 Oak Grove Dr}, \city{Pasadena}, \postcode{91109}, \state{CA}, \country{USA}}}

\affil[5]{\orgdiv{Department of Astronomy}, \orgname{California Institute of Technology}, 
\orgaddress{\street{1200 E.\ California Blvd.}, \city{Pasadena}, \postcode{91125}, \state{CA}, \country{USA}}}


\abstract{
Planetary debris disks around other stars are analogous to the Asteroid and Kuiper belts in the 
Solar System. Their structure reveals the configuration of small bodies and provides hints for 
the presence of planets. The nearby star Fomalhaut hosts one of the most prominent debris disks, 
resolved by HST, Spitzer, Herschel, and ALMA. Images of this system at mid-infrared wavelengths 
using JWST/MIRI not only show the narrow Kuiper-Belt-analog outer ring, but also that (1) what 
was thought from indirect evidence to be an asteroid-analog structure is instead broad, extending 
outward into the outer system; (2) there is an intermediate belt, probably shepherded by an unseen 
planet. The newly discovered belt is demarcated by an inner gap, located at $\sim 78$ au, and it 
is misaligned relative to the outer belt. The previously known collisionally generated dust cloud, 
Fomalhaut b, could have originated from this belt, suggesting increased dynamical stirring and 
collision rates there. We also discovered a large dust cloud within the outer ring, possible 
evidence of another dust-creating collision. Taken together with previous observations, Fomalhaut 
appears to be the site of a complex and possibly dynamically active planetary system.}




\maketitle

\section{Introduction}\label{sec1} 

The Infrared Astronomical Satellite (IRAS) found indications of planetary systems orbiting other stars, 
in the early 1980s. It was a great surprise that a handful of stars, amongst them
Fomalhaut, emitted far infrared radiation well in excess of their photospheric spectra \citep{aumann84,gillett86}: these infrared 
signals originated in circumstellar disks of dust and debris. Since then, the detection of 
thousands of exoplanets and hundreds of debris disks has increased our drive to understand the evolutionary 
connection between the planets and debris belts (i.e., asteroidal and Kuiper) of the Solar System with those in extrasolar systems.

Two complementary approaches have advanced our insight to these exoplanetary systems. 
The first is to study individual assembled and massive planets, either indirectly (e.g. transits and radial velocity 
oscillations) or by direct imaging. Indirect detections are most efficient close-in, 
while direct detections are more effective far-out. 
Transit and radial velocity studies of planet properties are biased against planets 
even at the Earth-Sun separation, while direct imaging can only reach a 
bit below the mass of Jupiter at current limits. The second approach, the study of planetary debris disks, focuses on the 
smallest bodies in a system, the dust and pebbles. Limited by resolving power and/or sensitivity 
at longer wavelengths and contrast near the central host star at optical/near-IR wavelengths, these
studies initially explored the outer zones of the systems. As more powerful instruments have 
been employed, debris disk studies have increasingly explored inward, and the structures of 
these disks have the potential to identify moderate-mass planets not identifiable by other means.  

Nonetheless, our knowledge of the inner zones of debris disks has been limited: (1) groundbased optical/near-IR
telescopes have poor sensitivity to the low surface brightness extended emission of debris disks; 
while (2) until now, cryogenic space infrared telescopes have had limited spatial resolution and 
cannot cleanly separate the disk components. JWST overcomes both of these challenges.

In this paper we report deep imaging observations of the debris system around Fomalhaut with JWST/MIRI 
and the Hubble Space Telescope (HST).
A forthcoming paper by Ygouf et al.\ discusses a deep search for exoplanets in this system using data from NIRCam on JWST.
This system brings forth the premier opportunity to resolve many of the questions 
about debris disks. At $7.66\pm0.04$ pc \citep{vanleeuwen07}, it is one of the nearest 
systems, allowing excellent spatial resolution \citep[e.g.,][]{macgregor17}. The high stellar luminosity 
(16.63 L$_{\odot}$) lights up its debris system \citep[L$_{\rm IR}$/L$_{\ast}$ = $8\times10^{-5}$;][]{rhee07} 
and also results in a large physical scale for belts of given temperature.
The system was known to host at least two spatially separate components, much like the Solar System, at similar thermal locations 
(i.e., at locations where the distance from the star compensates for its luminosity, so there is a 
similar level of irradiance). The outer Kuiper-belt analog (KBA) component of its disk system (at $\sim$ 140 au, 
a thermal location of $\sim 50-60~{\rm K}$) has been imaged in scattered light with HST \citep{kalas05,galicher13,gaspar20}, 
in the thermal infrared with the Spitzer Space Telescope \citep{stapelfeldt04} and the Herschel Space 
Observatory \citep{acke12}, and at radio wavelengths with SCUBA/JCMT and 
ALMA \citep[e.g.,][]{holland98,holland03,boley12,white17,macgregor17}. These observations have 
constrained this narrow ring to lie between 136 and 150 au, with a modest eccentricity of 
$e=0.12$. There is a faint halo of emission outside this belt \citep{kalas05,espinoza11,acke12} due to small grains on highly elliptical orbits or 
being blown out by photon pressure. An inner warm debris component, an asteroid belt analog (hereafter ABA)
at $\sim 150-170$ K \citep{lebreton13,su13,adams18} has also been inferred from the spectral energy distribution 
(SED), although it has previously not been spatially resolved. 
The system also exhibits a hot excess originating very close to the star \citep{absil09,mennesson13}.
Finally, Fomalhaut was identified long ago as a favorable venue for locating exoplanets 
\citep{roman59}, a suggestion supported by arguments that planets shepherd 
the debris system \citep[e.g.,][]{quillen06,chiang09,boley12}. A planet candidate
was identified by \cite{kalas08}, but is more likely to be
a dust cloud produced in a massive planetesimal collision \citep{absil09,lawler15,gaspar20}.

\section{Results}\label{sec2}

\begin{figure*}[!ht]
    \centering
    \figuretitle{The Architecture of the Fomalhaut Debris Disk System}
    \includegraphics[width=0.99\textwidth]{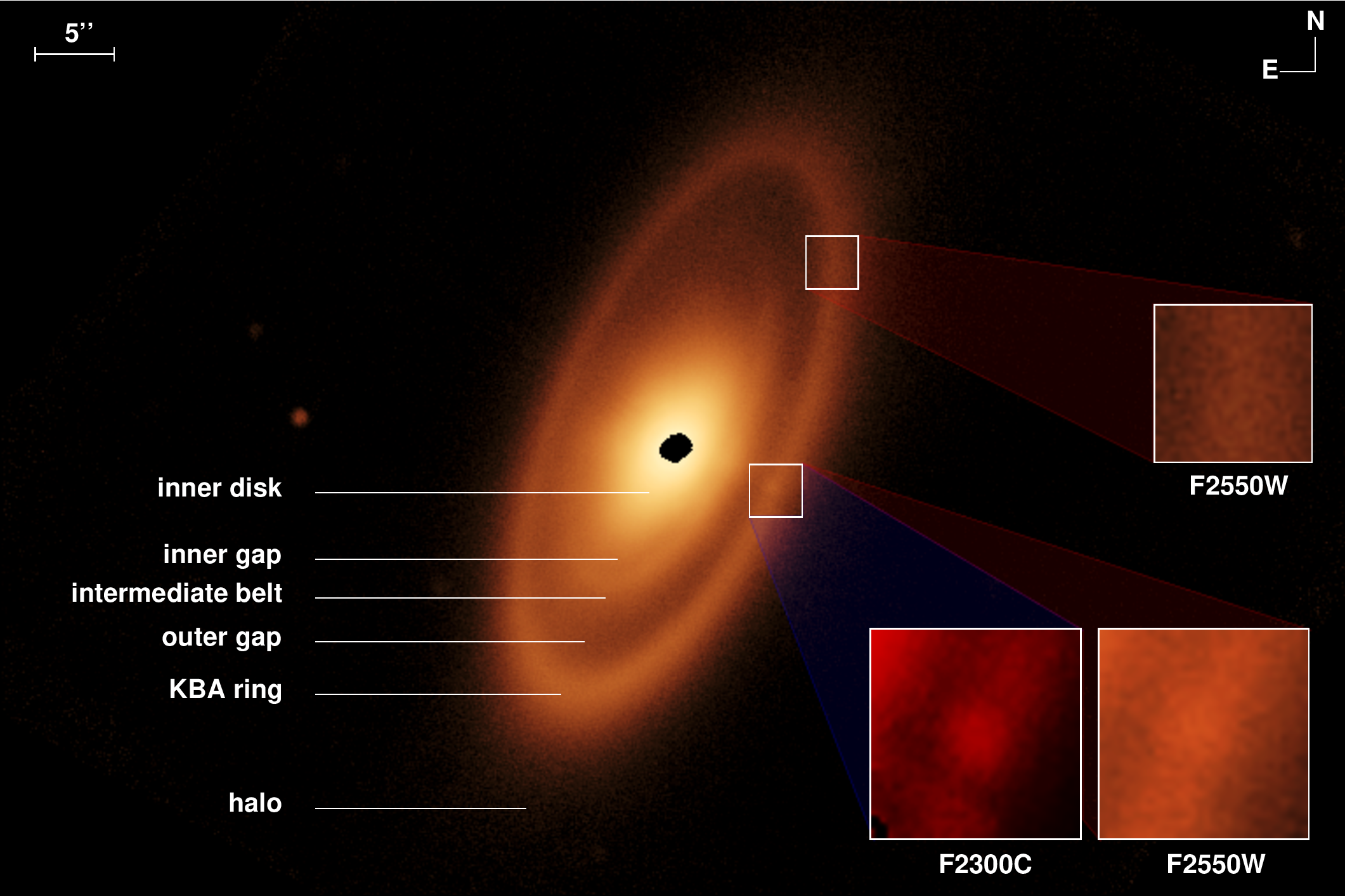}
    \caption{We highlight the various features identified in the JWST/MIRI observations of the Fomalhaut
    system in this 25.5 $\micron$ image. The lower two postage stamps enlarge the ``Great Dust Cloud'' that we identify 
    at 23.0 and 25.5 $\micron$, while the upper single postage stamp shows the projected location of Fomalhaut b 
    at the time of the observations (centered on the unbound trajectory solution; see Section ``Fomalhaut b''). 
    The 23.0 $\micron$ observations did not image this area. Fomalhaut b is not visible in the MIRI imaging.}
    \label{fig:labels}
\end{figure*}

We present observations taken with the Mid-Infrared Instrument (MIRI) onboard JWST as well as Cycle 27 
(2020-2021) coronagraphic imaging with the Space Telescope Imaging Spectrograph (STIS) onboard HST. 
Each of these instruments and modes probes different regions and dust sizes, 
providing a more complete picture of the Fomalhaut planetary system than was previously 
available. Below, we analyze the features revealed in our JWST observations, 
as highlighted in Figure \ref{fig:labels}, showcasing the 25.5 $\micron$ JWST image.
These include an extended inner disk, an inner gap, an intermediate belt, and a structure within
the outer KBA ring we call the ``Great Dust Cloud''.

\subsection{The Asteroid-belt analog}
\label{sec:ABA}

A variety of indirect arguments, based on SEDs and spectra, 
have identified a relatively warm component to debris disks that appears to be separate 
from the cold outer KBA rings \citep[e.g.,][]{Morales11}. The Fomalhaut system 
shares this behavior. 
\cite{su13} found that the SED of the ABA region has a fractional luminosity of $\sim$ $2\times10^{-5}$ and is 
well represented by a blackbody of 170 K, suggesting a radial distance of 11 au from 
the star. This suggests an enticing analogy to the structure of the
Solar System, with a cold Kuiper Belt and a warm Asteroid Belt, with the placement 
of the belts loosely associated with ice lines in the forming system \citep{su14}. Such 
structures remain by far the most popular explanation for the warmer component of debris disk 
SEDs in general \citep{ballering14,kennedy14,ballering17,geiler17}. However, inferences about debris disk
structure from SEDs can be highly degenerate between the assumed dust grain
absorption and emission properties, their distances from the star, scattering phase functions, porosity, 
sizes, and size-distributions. 

\begin{figure*}[!ht]
    \centering
    \figuretitle{The JWST Gallery of the Fomalhaut System with Deprojections}
    \includegraphics[width=1.0\textwidth]{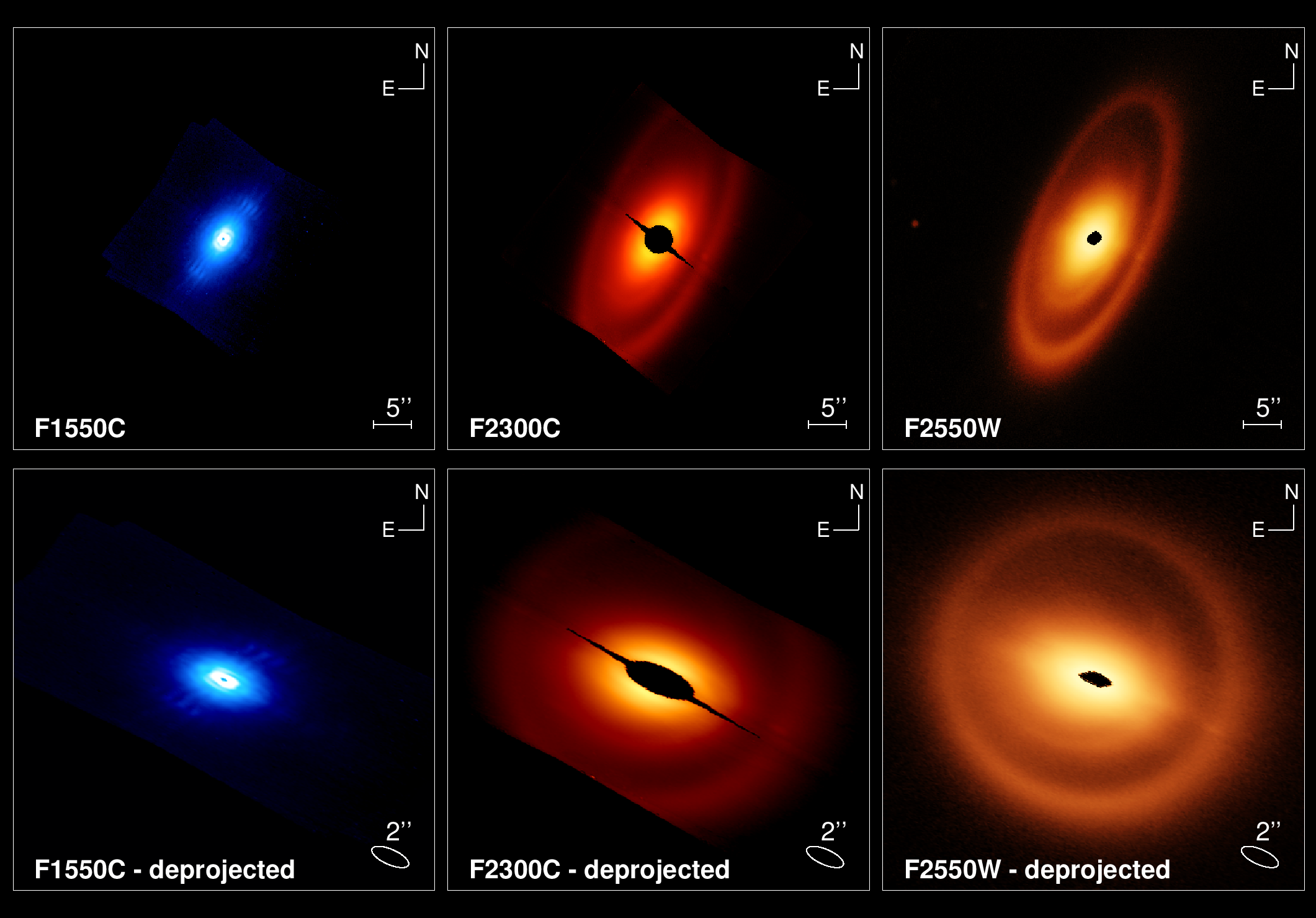}
    \caption{JWST observations of the Fomalhaut system at 15.5, 23.0, and 25.5 $\micron$. 
    The top row shows the system at its natural viewing
    angle, while the bottom row shows the observations deprojected by the best fitting
    inclination angle. The image scalings, projection fittings, and image reduction steps
    are discussed in Supplementary Information sections 1 and 2.}
    \label{fig:allJWST}
\end{figure*}

The 15.5, 23.0, and 25.5 $\micron$ JWST/MIRI images shown in Figure \ref{fig:allJWST} reveal a
disk system that is quite different from previous estimates based on photometry/spectrum 
fitting alone. The 25.5 $\micron$ observations spatially resolve an inner disk structure 
extending from at least $\sim 1\farcs2$ ($\sim$ 10 au) outward to $10^{\prime\prime}$ - $12\farcs5$ 
(77 au - 96 au; periapsis and apoapsis, respectively), including the intermediate belt. 
The inner limit ($\sim$ 10 au) is a result of image saturation, but is at the previously estimated  
location for the ABA component \citep[$\sim$ 11 au;][]{su13}. However, the inner disk
is significantly more extended radially than was assumed purely from its SED. 
The extent of this complex inner structure is confirmed at 23 $\micron$ as well. 
The 15.5 $\micron$ image is less diagnostic; nevertheless, 
it shows spatially extended emission tracking well with its longer wavelength counterparts. The ``gap'' at 
$\sim 1^{\prime\prime}$ in the F1550C data is likely a phase mask PSF subtraction residual, as it is 
similar to structures seen in commissioning data 
\citep{boccaletti22}. Therefore, the 15.5 $\micron$ data is not used to define the boundaries of 
the inner disk. Both 23.0 and 25.5 $\micron$ data show a prominent inner gap (see Figure \ref{fig:labels} labels) 
between the inner disk and the intermediate belt at $\sim 7\farcs5$ (periapsis) - $\sim 14\farcs2$ (apoapsis), 
which we discuss in more detail in Section ``The intermediate belt''. 

\begin{figure}
    \centering
    \figuretitle{Total Encircled Energy}
   \includegraphics[width=0.5\textwidth]{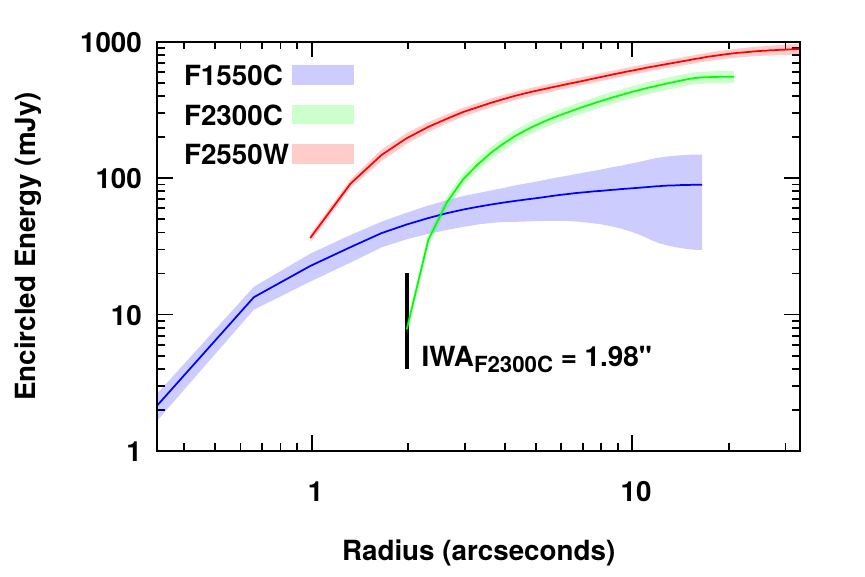}
    \caption{The measured total encircled energy at 15.5, 23.0, and 25.5 $\micron$ in the 
    observed orientation of the system, as a function of radius. Statistical errors were
    negligible. The dominant systematic errors are based on PSF subtraction scalings
    with a conservative 5\% and 10\% additional error added in quadrature at 25.5 and 23.0 $\micron$, respectively.}
    \label{fig:Noproject_fluxes}
\end{figure}

\begin{figure*}
    \centering
    \figuretitle{The Surface Brightness and Encircled Energy in the Deprojected Images.}
    \includegraphics[width=0.48\textwidth]{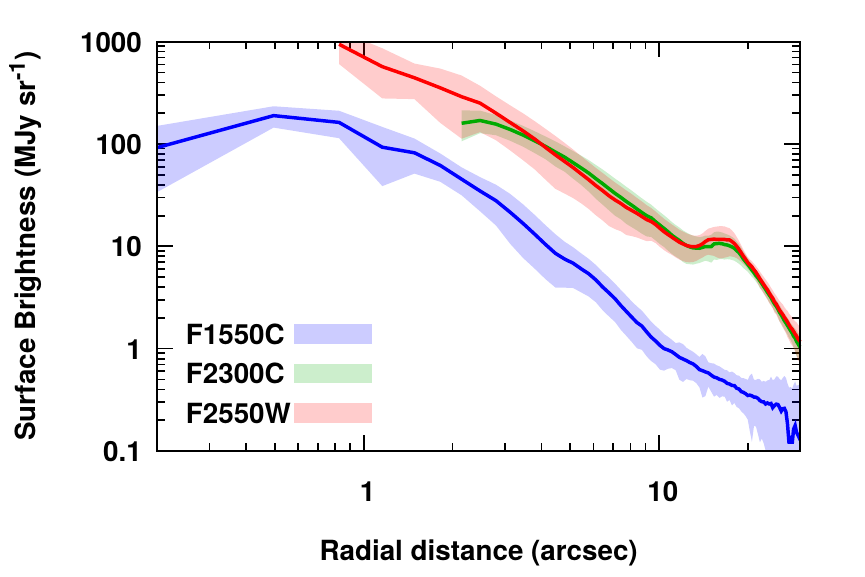}
    \includegraphics[width=0.48\textwidth]{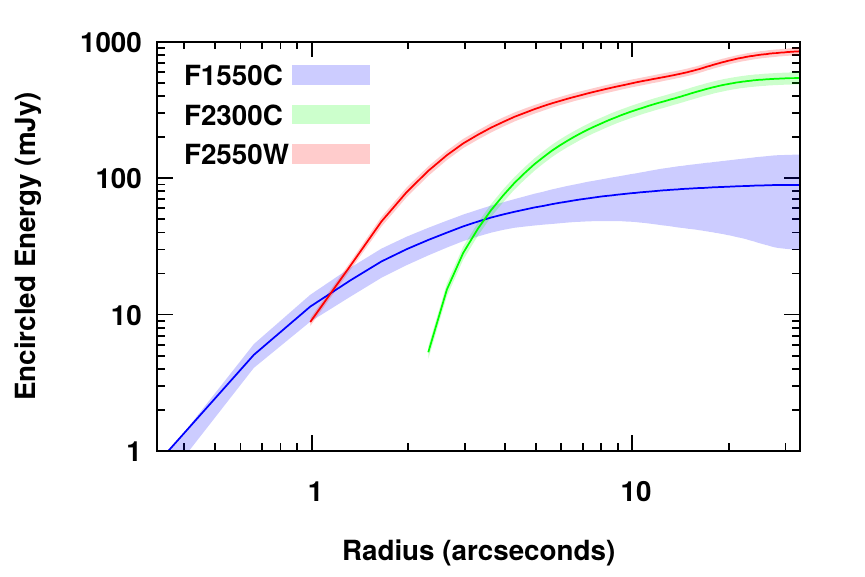}
    \caption{{\it Left panel:} The median surface brightness of the Fomalhaut disk as measured 
    in the de-projected image shown in Figure \ref{fig:allJWST} with pixel standard deviation 
    errors. {\it Right panel:} The total encircled energy as a function of radial distance as measured 
    in the de-projected images. Measurements inward of $\sim 1^{\prime\prime}$ at 15.5 $\micron$
    are mostly of PSF subtraction residuals. Statistical errors were negligible. The dominant 
    systematic errors are based on PSF subtraction scalings with a conservative 5\% and 10\% additional 
    error added in quadrature at 25.5 and 23.0 $\micron$, respectively.}
    \label{fig:fluxes}
\end{figure*}

In Figure \ref{fig:Noproject_fluxes}, we show 
the encircled energy at the three observed wavelengths with error bars, in the observed
orientation of the system (images in the top row of Figure \ref{fig:allJWST}). Due to saturation at the stellar core at 25.5 $\micron$ and to 
the Lyot mask at 23 $\micron$, the flux from the very inner regions is not included 
in these measurements, although at 25.5 $\micron$ we saturate only within 
$\sim 1\farcs2$ ($\sim$ 10 au). We used the position angle (PA) and inclination of the KBA ring, as determined by the MCMC orbit 
fitting at 25.5 $\micron$ (${\rm PA}=336\fdg28$, $\iota=67\fdg52$; see Supplementary Information section 2.2), 
to de-project the images at each observed wavelength (also shown in Figure \ref{fig:allJWST}). The de-projection
conserved flux; therefore, this view can be used to analyze surface structures as
well as to perform photometry on the disk components. In Figure \ref{fig:fluxes},
we show the median surface brightness and the encircled energy of the system at the 
observed wavelengths, as measured on the de-projected images. The calibration
of the 23.0 $\micron$ data (discussed in the Supplementary Information section 1.2) 
places it evenly on top of the 25.5 $\micron$ data in the outer regions of the system. The total flux of the 
inner disk -- including the intermediate-belt up to $13\farcs0$ -- is 
78$\pm$35 mJy, 375$\pm$38 mJy, and 560$\pm$36 mJy at 15.5, 23.0, and 25.5 $\micron$, 
respectively (measured from inner apertures of $0\farcs5$, $2\farcs3$, and $1\farcs0$ in radius, respectively).
The spatially extended nature of the inner disk, as resolved at 23.0 and 25.5 $\micron$, explains the low surface
brightness and weak detection at 15.5 $\micron$ and also contributes to the non-detection of the inner disk 
with HST/STIS (see Methods and Supplementary Information section 1.3 ``Cycle 27 HST Observations''). 

\begin{figure}[!t]
    \centering
    \figuretitle{The Recalibrated IRS Spectra and JWST Photometry of the Fomalhaut Debris Disk System.}
    \includegraphics[width=0.47\textwidth]{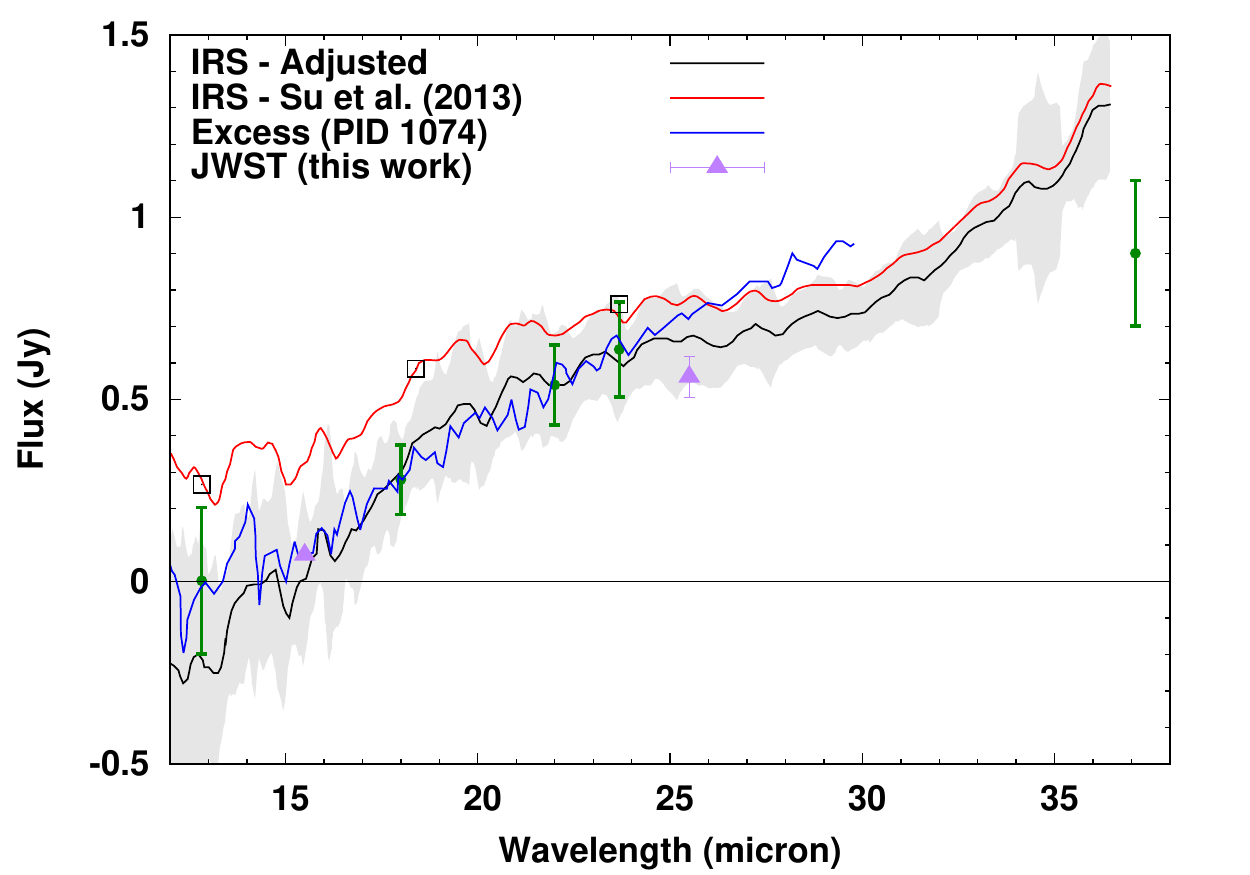}
    \caption{The black curve is the 
    spectrum with the calibration adjusted as described in Supplementary Information section 2.5;  
    the green points with 1$\sigma$ error bars show photometry 
    at 12.8, 18, 22, 24, and 37 $\mu$m (see the Supplementary Information for more information).  With light gray we plot the 
    1$\sigma$ errors in the spectrum, including an allowance for uncertainty in the flux of the stellar photosphere (see text). 
    The red curve shows the same data with the previous calibration discussed by 
    \cite{su13} and the open boxes are the averages of this spectrum over the photometric bandwidths (1 sigma SD 
    errors in the mean averages are similar to the sizes of the boxes). The blue line is the excess spectrum derived 
    from the data obtained under Spitzer PID 1074. The purple triangles with 1$\sigma$ error bars are from the images 
    at 15.5 and 25.5 $\mu$m.}
    \label{fig:IRSspectrum}
\end{figure}

The total flux measured at 15.5 $\mu$m in the inner region ($78\pm35$ mJy) is less than previously estimated 
from the Spitzer Infrared Spectrograph (IRS) spectrum \citep[320 mJy with an estimated error per resolution element 
of 190 mJy, or $<$ 60 mJy smoothed to photometric resolution;][]{su13}. We resolve this discrepancy by reviewing 
the calibration of the IRS data (see details in Supplementary Information section 2.5). The total disk flux, in the recalibrated spectra, is 
$102\pm60$ mJy at 15.5 $\micron$, and in good agreement with the measurement of $78\pm35$ mJy (Figure \ref{fig:IRSspectrum}).

We also can compare the flux from the 25.5 $\mu$m image with the recalibrated IRS spectrum. To match 
the IRS slit dimensions and orientation, we measure the flux in a 12$^{\prime\prime}$ wide 
(and 1$^{\prime}$ long) box, set at a PA of 64$^{\circ}$ \citep{stapelfeldt04} in the F2550W 
image and assign a 10\% error. The flux measured within the IRS slit area (561$\pm$56 mJy) 
lies within the errors, although at a lower flux, than the recalibrated IRS spectrum value at 25.5 $\micron$ (Figure \ref{fig:IRSspectrum}). 
The minor discrepancy is likely to indicate a small amount additional flux generated 
inside a radius of $1\farcs2$ (10 au), which we are missing due to saturation. 
The small amount of missing flux shows that the disk cannot extend far inside 10 au, 
nor contain large numbers of small grains ($\sim$ 1 $\mu$m), or its emission in the 10 $-$ 17 $\mu$m 
range would exceed the limits imposed by the spectrum. In addition,  the slope of the surface brightness profile
at 25.5 $\micron$ in Figure \ref{fig:fluxes} may be decreasing inwards of 2$^{\prime\prime}$ compared with further out. 
These constraints suggest an inner edge in the 7 - 8 au range.

Many debris systems have now been imaged with large ground-based telescopes in scattered light at near-IR wavelengths  \citep[e.g.,][]{esposito20}. 
At the high resolution of these images, inner dust structures would be readily detected. However, despite the 
emission from these regions in the 15 - 25 $\mu$m range, virtually all of them have inner holes in the near-IR scattered light 
\citep[to the limits achieved,][]{esposito20}. Attempts to image inner structures in optical scattered light with 
coronagraphs on HST have also been unsuccessful \citep[e.g.,][this paper]{wolff23}, nor have they been detected with ALMA \citep[e.g.,][]{su16}. 
The low levels of scattered light in the inner Fomalhaut system appear to be a common characteristic. 
This behavior perhaps arises from minimum grain sizes larger than previously assumed 
 and the resulting low scattering efficiency at near normal incident angles \citep[e.g.,][]{wolff23}.

Our observations show that the Fomalhaut inner disk system is quite different from that of our
own Solar System. While the Asteroid belt is relatively narrow (located between $\sim$ 2.1 and 3.3 au),
shepherded by a massive planet, the inner region of Fomalhaut hosts an extended disk component.
Which architecture is more common? Our upcoming observations of 
the Vega and $\epsilon$ Eridani system and other JWST observations of nearby debris disks should help to steer the discussions on this subject. 

\subsection{The intermediate belt}
\label{sec:innerbelt}

Perhaps our most intriguing result is the detection of an intermediate belt in the Fomalhaut system. The belt 
is clearly resolved towards both apices although not nearly as prominently towards the forward direction (N direction as viewed 
from Earth) of the disk as the back side (S direction as viewed from Earth). This is also true for the outer KBA ring, 
which is brighter on the southern arc than the northern in thermal imaging. This is opposite to how the system appears 
in scattered-light HST images, where efficient forward scattering brightens the northern arc.

We fit the inner and outer 
boundaries of the feature with a Keplerian orbit on the de-projected F2550W image (Figure \ref{fig:allJWST}). The outer
boundary was the easier one to trace of the two, and has a semi-major axis of $\approx 104$ au and an eccentricity of e=0.265.
Interestingly, the fitted orbit was inclined by $\iota\approx22\fdg9$ with respect to the de-projected plane. This agrees
with the contour fits (see Supplementary Figure 10), pointing towards an inclination offset between the inner and outer regions.
The inner boundary of the feature can be fit with an elliptical path with a semi-major axis of $\approx 83$ au and 
eccentricity of 0.31. The fitted relative inclination of the inner boundary is much smaller at $\iota\approx 7\fdg4$.
The width of the belt varies from ($\sim$ 7 to 20 au). The peak of the belt is at $\sim 16.8$ MJy sr$^{-1}$
and is $\sim 1.5$ MJy sr$^{-1}$ above the value of the average background interior and exterior of it.

The feature is demarcated externally by the previously well resolved cleared region, already identified in HST images, 
and an inner gap in the system, imaged by the JWST/MIRI observations. The inner gap is $\sim 1\farcs3$ (10 au) wide at 
$\sim 10^{\prime\prime}$ ($\sim$ 78 au at center) separation from the central star. The gap is clearly present at both 
23.0 and 25.5 $\micron$ and is therefore a real feature of the system, likely 
a product of gravitational perturbations from an extrasolar planet. An alternate theory for its presence could be 
intense dust production at the location of the intermediate belt, resulting for example from the shearing of a 
recently produced large dust cloud. Detailed numerical modeling of this feature and the characteristics of the 
shepherding planet will be presented in a follow-up paper; preliminary models using DiskDyn do place upper limits on the 
mass of the perturbing planet around 1 M$_{\rm Saturn}$, which is significantly lower than the detection limits 
of previous imaging surveys \citep[e.g.,][]{gauchet16,maire14,kenworthy13}.

\subsection{The Great Dust Cloud}
\label{sec:GDC}

An intriguing feature present in both the 23.0 and 25.5 $\micron$ observations is a round cloud in the KBA ring,
highlighted in Figure \ref{fig:labels}, which we call the ``Great Dust Cloud'' (GDC). Since it appears at both wavelengths, 
it cannot be a reduction artifact. We estimate its brightness to be 61$\pm$12 $\mu$Jy at 23.0 and 
89$\pm$13 $\mu$Jy at 25.5 $\micron$ in excess of the belt brightness, within an $0\farcs5$ radius aperture. These flux ratios
hint at the source being cold dust. Assuming a standard -3.65 size distribution slope with a minimum grain size of 1 $\micron$, we estimate an emission surface 
area (4$\pi r^2$) of $\approx 6\times10^{19}~{\rm m}^2$ for the dust within the cloud (mass of $1.36\times10^{-7}$ M$_{\rm Earth}$ 
up to 1 mm radius grains assuming 2.7 g cm$^{-3}$ bulk density). This is an order of magnitude more dust than we estimated 
for the scattering surface for Fomalhaut b \citep{gaspar20}. A catastrophic collision between two objects of 
355 km radius at a velocity of 360 m s$^{-1}$ (which corresponds to $\sim$10\% of the orbital velocity at the GDC's location) 
would be able to produce the amount of dust observed, based on collisional calculations using equations in \cite{gaspar12a}.
Alternatively, chance alignment with a background source is not ruled out by this single epoch observation.

\subsection{Fomalhaut b}
\label{sec:Fomb}

Fomalhaut b, along with the planets of HR 8799, were the first directly imaged extrasolar planet 
candidates orbiting a star \citep{kalas08,marois08}, although the status of Fomalhaut b was always contentious due 
to non-detection at near-infrared wavelengths \citep{kalas08,janson12}. There have been various hypotheses explaining the nature 
of the object, including scattered light from a circumplanetary disk \citep{kennedy11,kenyon15} or a giant impact between two larger 
planetesimals \citep{galicher13,kenyon14,lawler15}.
\cite{gaspar20} showed that the HST observations reveal the object to have faded over the decade of observations
as well as broadened in size, indicative of an expanding dust cloud. The object is moving on an escape trajectory,  
pointing to these dust particles being small in size with motions dominated by stellar radiation pressure, indicating they were 
created in a recent massive collision.

Based on models presented in \cite{gaspar20}, the object should be at $\Delta\alpha \approx -9\farcs809$ and
$\Delta\delta \approx 11\farcs665$ on its unbound orbit (or at $-9\farcs377$, $11\farcs144$ if on a bound orbit)
at the time of the JWST observations. We estimate the flux from this object using the best fitting expanding dust cloud 
model evolved to our observing date \citep{gaspar20}. The estimated surface brightness is $\approx 0.5$ MJy sr$^{-1}$ 
at 25.5 $\micron$, which is well within the 1$\sigma$ brightness variation at a radial distance of $\sim 15\farcs2$ (see 
Supplementary Figure 11). Consistent with this estimate, we do not detect any observable signature at either of the coordinates in the imaging
at 25.5 $\micron$ (the region was not imaged at the other wavelengths). 

A lingering concern for the collisional hypothesis was the lack of disk material observed interior to the 
KBA ring, indicating a quiescent and low planetesimal density at the point of origin for Fomalhaut b. 
Our JWST observations, however, reveal that this location -- the intermediate belt --  has high dust density. 
The asymmetric nature of the belt also hints at increased collisional activity in the region. The JWST observations
therefore address the concerns about the expanding cloud hypothesis. 

\section{Discussion}

The cold KBA debris belts are often well resolved, particularly with ALMA \citep[e.g.,][]{macgregor13,ricci15,macgregor17,faramaz21}. 
Planetary systems will likely lie within these belts. The gravitational fields of planets will carve 
structures in any interior debris components; the placement of these components  gives clues to the 
existence and architecture of exoplanetary systems. Until now, these clues have not been  accessible: 
(1) infrared SEDs indicate the presence of interior disk components but cannot reveal their structure well; and (2) 
imaging scattered light in the visible and submm emission with ALMA  have not even achieved solid detections in these regions. 

This problem has been abolished with MIRI on JWST, which combines the necessary resolution, sensitivity 
to low surface brightness, and an array of useful tools (multiple imaging filters, coronagraphs, integral 
field unit spectroscopy). We have turned this revolutionary capability onto the best-resolved bright debris 
system, that around Fomalhaut. The results have not disappointed. We have detected the dust being blown 
out of the KBA ring, and in addition have found (1) a narrow intermediate debris belt interior to the KBA ring, (2) a diffuse 
distribution of dust in a disk-like configuration closer to the star, and (3) a possible dust cloud located
within the KBA ring itself (this source being a background object is not currently ruled out). 
The intermediate belt also reveals increased dust production and 
collisional activity at the location assumed for the origin of the object Fomalhaut b, supporting  
the hypothesis that it was produced in a massive collision. 

In addition, the structure of the inner debris indicates that it is shepherded through the gravitational influence of 
yet unseen planets. Planets as low in mass as Neptune are sufficient to carve the inner belts. That is, our observations 
provide insights to the Fomalhaut planetary system extending down to common planet masses. The structures of 
the massive debris belts, their alignment offsets, and the indications of massive collisional events (Fomalhaut b and the GDC)
all highlight that the 440 Myr old star Fomalhaut is surrounded by a complex  planetary system undergoing dynamical perturbations. 

\section{Methods}

We describe the JWST/MIRI observations and detail the reduction steps of the 25.5 $\micron$ F2550W dataset in this 
section. Similar information about the coronagraphic 15.5 and 23.0 $\micron$ JWST/MIRI and optical
wavelength HST/STIS observations can be found in Supplementary Information section 1.
We show our final reduced and processed F2550W image of the Fomalhaut system in Supplementary Figure 1. 

The MIRI wavelengths we observed (15.5, 23.0, and 25.5 $\micron$) equate to the peaks of blackbodies 
with temperatures of 186, 125, and 113 K, respectively. Assuming 10 $\micron$ radius classic astro-silicates 
\citep{draine84}, these temperatures correspond to distances of 7.1 au ($0\farcs9$), 15.7 au ($2\farcs$0), and 
19.3 au ($2\farcs5$) at Fomalhaut, respectively. The thermally equivalent distances in the Solar System are 
1.8, 3.9, and 5.0 au, i.e. regions extending from the orbit of Mars to that of Jupiter. These angular separations are 
readily resolved at 15.5 $\micron$ with the 4QPM and at 25.5 $\micron$ via non-coronagraphic imaging with MIRI. 
In addition, colder ($\sim 50$ K) dust, located for example in the KBA regions, will still emit $\approx 20\%$ of its peak 
flux at 25.5 $\micron$ and was therefore also expected to be detected in the observations.

\subsection{JWST Observations}

Our Cycle 1 GTO observations (PID 1193) were carried out on October 22, 2022, for both MIRI 
(Wright et al., in press) and NIRCam \citep{riekem23}, using a continuous
non-interruptable sequence. The observing program is a result of a collaboration between 
the MIRI and NIRCam GTO teams, who joined efforts on the same targets 
(Fomalhaut, Vega, and $\epsilon$ Eridani). The MIRI program focused on observations at longer wavelengths to reveal 
the inner disk structure, while the NIRCam observations were designed to detect extrasolar 
planets.  This paper discusses the MIRI observations only. Without prior knowledge of the structure of the
ABA disk, we designed the observations to probe various regions in the system. We therefore included high-contrast imaging via
coronagraphy at 15.5 and 23.0 $\micron$ and reference differential imaging (RDI) via 
non-coronagraphic imaging at 25.5 $\micron$. The latter took advantage of the remarkably stable telescope Point Spread 
Function (PSF). 

Reference PSF observations were taken at each wavelength 
contemporaneously with the science target data acquisition as well as background measurements 
for the MIRI coronagraphic observations. Apart from a single guide star failure for our MIRI 
F2550W imaging visit to the PSF reference source, which was repeated on November 13, 2022, our
imaging sequence was successfully executed. We summarize the complete set of JWST/MIRI observations 
in Supplementary Table 1., in the sequence they were carried out. 
The MIRI+NIRCam sequence was optimized for various aspects of our observing goals and also to 
save observatory (slewing) overheads. The MIRI sequence was specifically designed to provide long downtimes between using the same 
detector locations to avoid latent images. For PSF reference, we observed 19 PsA (M3III) 
at all wavelengths, which proved to be an excellent choice. Although 19 PsA is 100x fainter 
than Fomalhaut in the V band, it is 10-15\% brighter than Fomalhaut at the MIRI wavelengths and its 
photospheric emission across the MIRI bands  mostly follows a Rayleigh-Jeans function, providing a close match to the spectrum of Fomalhaut. 19 PsA is also located at an 
offset of only 3.3$^{\circ}$, providing similar background characteristics as well as ensuring 
very little thermal variation in the telescope optics. 

The target observations were obtained at two rotation angles, offset by 10 degrees, to provide rotational 
coverage and dithering. The coronagraphic PSF observations were obtained on a 9pt small grid dither (SGD) pattern,
which ensured a well-matched PSF. The SGD is especially important for the four quadrant 
phase mask (4QPM) observations at 15.5 $\micron$ (and as we learned less important for the 
Lyot at 23.0 $\micron$; Supplementary Information section 1.2). Having a brightness matched PSF reference ensured 
similar noise characteristics as for the target, which was especially important for the F1550C observations, 
where only the two closest matching dither position observations of the 9 point small-grid-dither 
pattern were used in our classical reference differential imaging reductions. The 
non-coronagraphic observations 
with the F2550W filter were also spatially dithered in a larger ``extended'' source pattern 
at 4 pointing positions to correct for latent images and uneven background subtractions. Additional 
details about the individual observations can be found in the following subsections and in Supplementary
Information section 1. 

\subsection{MIRI 25.5 $\mu$m imaging and processing}

Fomalhaut itself is quite bright even at these wavelengths ($\sim 2.52$ Jy), therefore we had to mitigate 
the effects of saturation and latent images. While the standard full imager integration time is 2.775 
seconds per group (aka frame), shorter integrations are available in subarray modes. We used the 
Brightsky subarray for the non-coronagraphic imaging program, which reads out a smaller 512$\times$512 array with only 
0.865 s integrations. These shorter frame times ensured that only a handful of pixels saturated 
in the stellar core, while we were still able to record 5 frames per integration. The 56$^{\prime\prime}$ field 
of view (FOV) of the Brightsky subarray also ensured that the entire Fomalhaut disk was within the imaged area, 
even using the larger extended source dither pattern, at any possible orientation. The extended source 
dither pattern enabled us to construct an even background and further mitigated the effects 
of imager saturation and detector/background artifacts.

The data were downloaded from the Mikulski Archive for Space 
Telescopes (MAST), which archives the observations at all processing stages.  
For the reductions of the F2550W  images, we used the raw uncalibrated images (\texttt{\_uncal.fits}) 
and reduced them with the pipeline (v.\ 1.8.2) to stage 2. In general, these steps involve 
detector-level corrections, 
ramp-fitting, observing-mode corrections, and calibrations. We turned off dark corrections for the 
reduction steps, as the background images we later generated using the reference PSF observations 
included the same dark patterns, given we employed an identical observing sequence for the reference 
source. These stage 2 products (\texttt{\_cal.fits} files) still contain detector-level artifacts 
for MIRI, typically varying background levels, row and column effects, saturation columns and 
slewing tracks (in our particular case), and the ``tree-ring'' structure \citep{ressler08}. 
Further image reductions to remove these artifacts, as well as the image processing steps, were performed
using IRAF. The ``tree-ring'' was removed with a background correction image, produced by median combining 
the PSF images (19 PsA) without any alignment. Even with the extended dither pattern the stellar 
contribution was not removed completely with an initial median combination. We corrected this, 
by generating a median combined PSF, which was then removed from individual images prior to merging
them. This process was repeated iteratively, with the generated background also removed from the 
individual images at each step. The background image produced after five iterations was used to 
correct the Fomalhaut and the reference PSF images.
We removed the remaining row effects by median averaging the rows and subtracting these 
values from the images, while the constant background level was set to zero by an iterative 
3$\sigma$ clipped median sky subtraction. These steps were executed on both the target and 
reference observations at each dither position independently. We did not execute the stage 3
processing on the dataset; therefore, geometric distortion corrections were not applied to the images
we present in this work, but are estimated to be less than 1.5 px over the $\sim 18^{\prime\prime}$
extent of the Fomalhaut disk \citep[$<0.9\%$ over the entire detector;][]{bouchet15}.

Our Fomalhaut data are the first PSF reference subtracted high SNR observations of any object 
with JWST at these wavelengths, therefore no standard reduction steps have been developed. To determine
the ideal processing steps, we tested multiple methods with the dataset. As the initial PSF 
reference observations failed at 25.5 $\micron$, we first reduced the dataset using theoretical 
PSFs and a lower-quality noisier observed PSF ($\beta$ Dor; PID 1023) 
obtained during commissioning, and also via angular differential imaging (ADI) techniques, while waiting for the 
repeat PSF observations. The theoretical PSF, generated using \texttt{WebbPSF} \citep{perrin14} and 
employing the contemporaneous optical path difference (OPD) information of the observatory, 
proved to be insufficient for our purposes, revealing significant differences in scaling and patterns 
between the observed and modeled PSF structures. The variances are likely to arise predominantly due to 
the brighter-fatter-effect in the MIRI detectors (I. Argyriou et al. 2023, in preparation), which is not 
included in the \texttt{WebbPSF} simulations. Efforts to improve on the theoretical PSF proved to 
be fruitless. The PSF observed during commissioning provided a much better subtraction of the 
stellar core, but has a much lower SNR and therefore introduced noise levels greater than the disk 
signal we were seeking to reveal even near the core. 
Additionally, these observations used the SUB256 subarray, which has a smaller FOV of only 28$^{\prime\prime}$ 
(smaller than the full extent of the Fomalhaut KBA ring at 38$^{\prime\prime}$), thereby 
disqualifying this PSF to be used in reducing the entire FOV. 

However, some interesting discoveries were made about the JWST Optical Telescope Assembly (OTA) and MIRI optical system when we performed 
classical ADI reductions. The two Fomalhaut F2550W observations, offset by 10$^{\circ}$ in position 
angle, were separated by 10 hours, to allow the detector to erase latent artifacts 
without annealing. Assuming minor temporal variations in the optical system, as an initial attempt 
we combined all Fomalhaut target observations to produce a ``super'' PSF and used this as a reference 
to subtract. This method resulted in an image with expected self-subtraction residuals due to the 
extended disk signal, and only marginally better results within the PSF core than what we achieved 
using the theoretical PSF. As a second method, we subtracted only 
the matching dither position pairs from each other. These subtractions resulted in virtually
zero residuals of stellar contribution, even with the 10 hour gap between their acquisition (with the disk self-subtractions obviously still present). 
This result is true at all four dither positions. The MIRI PSF -- at least at these 
wavelengths -- is remarkably stable but position dependent (at the levels revealed by precise PSF subtraction).

The stability and position dependence of the PSFs were corroborated with the repeat observation of 
our PSF target at 25.5 $\micron$, which provided residual-free subtractions, as long as PSFs obtained 
at the same detector locations were used. To highlight the position dependence of the PSFs, we show 
their differences relative to each other (the same reference source - 19 PsA - subtracted from 
itself) in Supplementary Figure 2.
The residuals are on the level of the disk signal we are detecting. To demonstrate the 
position dependence of the PSFs and accurate matching when using the same position, 
we show the PSF subtraction residuals for Fomalhaut at the first dither position in Supplementary Figure 
3. When using the PSF obtained at the same detector position (D1), there are no 
detectable residuals, while the PSFs obtained at positions 2 and 3 yield noticeable patterns. 
The PSF imaged at D4 is better, but still shows some residuals. Subtractions of 
the target images taken at D2, D3, and D4 yielded similar results. The residuals are speckle-like
in pattern and are not a result of the uncorrected geometric distortions.

Although the PSF reference observations were executed three weeks after the target observations, 
due to their initial failure, the subtractions again showed no detectable stellar residuals, as 
long as the PSFs were subtracted in pairs corresponding to the same detector positions. 
Our PSF target, 19 PsA, is slightly brighter than Fomalhaut at 25.5 $\micron$ (2.88 vs.\ 2.52 Jy) 
and we obtained identical exposures for it as we did for Fomalhaut, thereby guaranteeing matching PSFs. 
We achieved residual-free subtractions using a PSF 
scaling factor of 0.9; at 0.87 and 0.93 under- and over-subtraction residuals were apparent. 
Following masking for latent patterns, individually for each of the eight images, we median 
combined the dataset with 3$\sigma$ clipping around the data median. 

We derived the absolute calibration from the reference star.  For MIRI, the minimum number of frames 
per ramp is 5 to ensure a well sampled integration slope. The first and last frames of a ramp are 
not used, therefore, the slopes only consisted 
of 3 frames for analysis purposes. The current pipeline does not apply a reset switch charge decay 
(RSCD) correction (Morrison et al., in prep) for such short ramps, therefore the absolute calibration 
of the 25.5 $\micron$ images was uncertain. To verify the absolute calibration, we used the 
observations of the PSF target, 19 PsA, which were carried out in an identical way to the 
target observations, therefore should be subject to the same systematic offsets. The reference observations were of a clean point source, allowing us to compare it to a theoretical
PSF. While the particular speckle pattern of the theoretical PSF is not an ideal match, as shown
previously, we can use it globally for photometric calibration. For this purpose, we used the 
theoretical PSF that we generated for our initial reductions,
using the available OPD maps of the telescope. While the reference observations were taken three
weeks after the target observations, we were informed by the Space Telescope Institute that 
the optical maps were identical for the purposes of modeling the 25.5 $\micron$ PSF (Marshall Perrin
and Matthew Lallo, private communication). This was corroborated by the clean PSF subtractions 
shown in Supplementary Figure 3. We confirmed that the core of the 
theoretical PSF agrees with the observations, by making a comparison to the $\beta$ Dor observations previously introduced. 
In Supplementary Figure 4, we show the radial profile of the observed PSFs and 
the fitted theoretical one, where we apply a radial location dependent scaling function to the
latter. The radial-dependent scaling function was necessary to obtain a better fit at radial offsets
greater than 10$^{\prime\prime}$, varying linearly from a value of 36.5 at the stellar core to 
13.33 at 20$^{\prime\prime}$. The total integrated flux of the scaled theoretical PSF is
2580 mJy, while the predicted photospheric brightness of 19 PsA was 2880 mJy. As 19 PsA is a small
amplitude (1-2\%) variable star, we also used the flux value determined from our fitting 
to the Fomalhaut observations, where we determined a scaling factor of 0.90 for it to fit
the target observations. Fomalhaut has a brightness of $2570\pm50$ mJy at 25.5 $\micron$, yielding
a brightness of $2855\pm55$ for 19 PsA, in excellent agreement. Based on these estimates, we applied a 1.11 ($\pm 0.02$)
scaling factor to the F2550W images for absolute calibration. 

\bmhead{Data Availability}

The final reduced and processed JWST MIRI imaging data that we present in this
paper (and ones including geometric distortion corrections) can be downloaded from 
\url{https://github.com/merope82/Fomalhaut}. The raw and general pipeline processed
data products will be available from the Mikulski Archive for Space Telescopes (MAST) 
at the Space Telescope Science Institute following their public release.
The JWST observations are associated with program \#1193, while the HST observations 
are associated with program 15905.

\bmhead{Acknowledgments}
We would like to acknowledge the assistance of our program coordinators, Crystal Mannfolk and 
Blair Porterfield, especially in ensuring a quick repeat of the initially failed PSF reference observations. We want
to thank Marshall Perrin and Matthew Lallo for updating us on the primary mirror alignment while we waited for the
repeat of the PSF observations. Financial support for this work was provided by Grants 80NSSC22K1293 
and HST-GO-15905.001-A to the University of Arizona. This work is based in part on observations made 
with the NASA/ESA/CSA JW and Hubble Space Telescope. The data were obtained from the Mikulski Archive 
for Space Telescopes at the Space Telescope Science Institute, which is operated by the Association of 
Universities for Research in Astronomy, Inc., under NASA contract NAS 5-03127 for JWST. 

\bmhead{Author Contributions} The observations were designed by A.G. and G.R., the data were
reduced and processed by A.G. and S.W., while all co-authors contributed to the data analysis
and writing of the manuscript.

\bmhead{Competing Interests}

The authors do not declare any competing interests.

\clearpage
\newpage
\onecolumn
\pagestyle{plain}

\renewcommand{\figurename}{Supplementary Figure} 
\renewcommand{\tablename}{Supplementary Table} 

\setcounter{section}{0}
\setcounter{subsection}{0}
\setcounter{figure}{0}
\setcounter{table}{0}

\begin{center}
\noindent{\Large\bf Supplementary information}
\end{center}


\section{Observations - continued}

\begin{table}[!h]
\begin{center}
\caption{JWST MIRI observations of the Fomalhaut system (Oct 22, 2022)\label{tab:JWSTobs}}
\begin{tabular}{lllllll}
\toprule
Target                      & Filter  & PA$^{\ddagger}$ & N$_{\rm group}$ & N$_{\rm int}$ & Dither               & Time (s) \\
\midrule
19 PsA (PSF Ref.)$^{\ast}$  & F2550W       & 59.20   & 5               & 52            & 4pt Ext$^{\dagger}$  & 1076.4  \\
19 PsA (PSF Ref.)           & F2300C       & 59.20   & 95              & 7             & 9pt SGD              & 1956.6  \\
PSF background              & F2300C       & 59.22   & 95              & 7             & None                 & 217.4   \\
19 PsA (PSF Ref.)           & F1550C       & 59.22   & 168             & 5             & 9pt SGD              & 1820.6  \\
PSF background              & F1550C       & 59.25   & 168             & 5             & None                 & 202.3   \\
Fomalhaut (Rot 1)           & F2550W       & 50.65   & 5               & 52            & 4pt Ext$^{\dagger}$  & 1076.4  \\
Fomalhaut (Rot 1)           & F2300C       & 50.65   & 102             & 29            & None                 & 967.5  \\
Fomalhaut background        & F2300C       & 56.02   & 102             & 29            & None                 & 967.5  \\
Fomalhaut (Rot 1)           & F1550C       & 52.06   & 211             & 18            & None                 & 914.4  \\
Fomalhaut background        & F1550C       & 56.07   & 211             & 18            & None                 & 914.4  \\
Fomalhaut (Rot 2)           & F1550C       & 62.05   & 211             & 18            & None                 & 914.4  \\
Fomalhaut (Rot 2)           & F2300C       & 60.63   & 102             & 29            & None                 & 967.5  \\
Fomalhaut (Rot 2)           & F2550W       & 60.63   & 5               & 52            & 4pt Ext$^{\dagger}$  & 1076.4 \\
\botrule
\end{tabular}
\end{center}
\begin{flushleft}
\footnotetext{The table reflects the observing sequence we employed to obtain the dataset, which was optimized to reduce latent images in
the MIRI detectors, providing long gaps between observations performed at the same detector location/wavelength.}
\footnotetext[{\ast}]{This observation failed in the initial sequence due to guide star failure and was repeated on Nov 13, 2022.}
\footnotetext[{\dagger}]{The Brightsky subarray 4 pt extended dither pattern used the \#6 starting set with a single set.}
\footnotetext[{\ddagger}]{Position angle of observed aperture.}
\end{flushleft}
\end{table}

\subsection{JWST/MIRI 25.5 $\micron$ imaging extra Figures and Tables}

\begin{figure*}[!h]
    \centering
    \figuretitle{The JWST/MIRI F2550W 25.5 $\micron$ non-coronagraphic image of the Fomalhaut debris disk system}
    \includegraphics[width=1.0\textwidth]{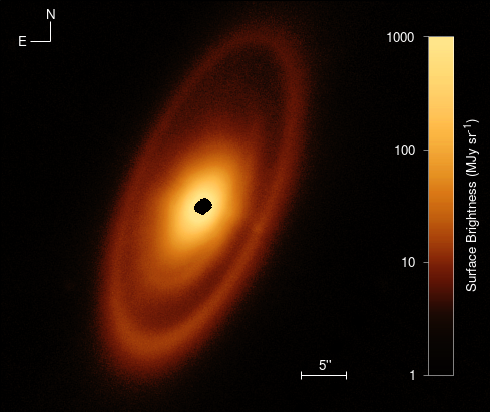}
    \caption{The image, produced via classic reference PSF subtraction, is displayed in logarithmic scaling 
    between -4.6 and 2000 MJy sr$^{-1}$, with the colorbar showing (part of) the mapping.}
    \label{fig:F2550W}
\end{figure*}

\begin{figure*}[!h]
    \centering
    \figuretitle{Variations of the reference PSF (19 PsA) as a function of detector position in 
    the F2550W images.}
    \includegraphics[width=1.0\textwidth]{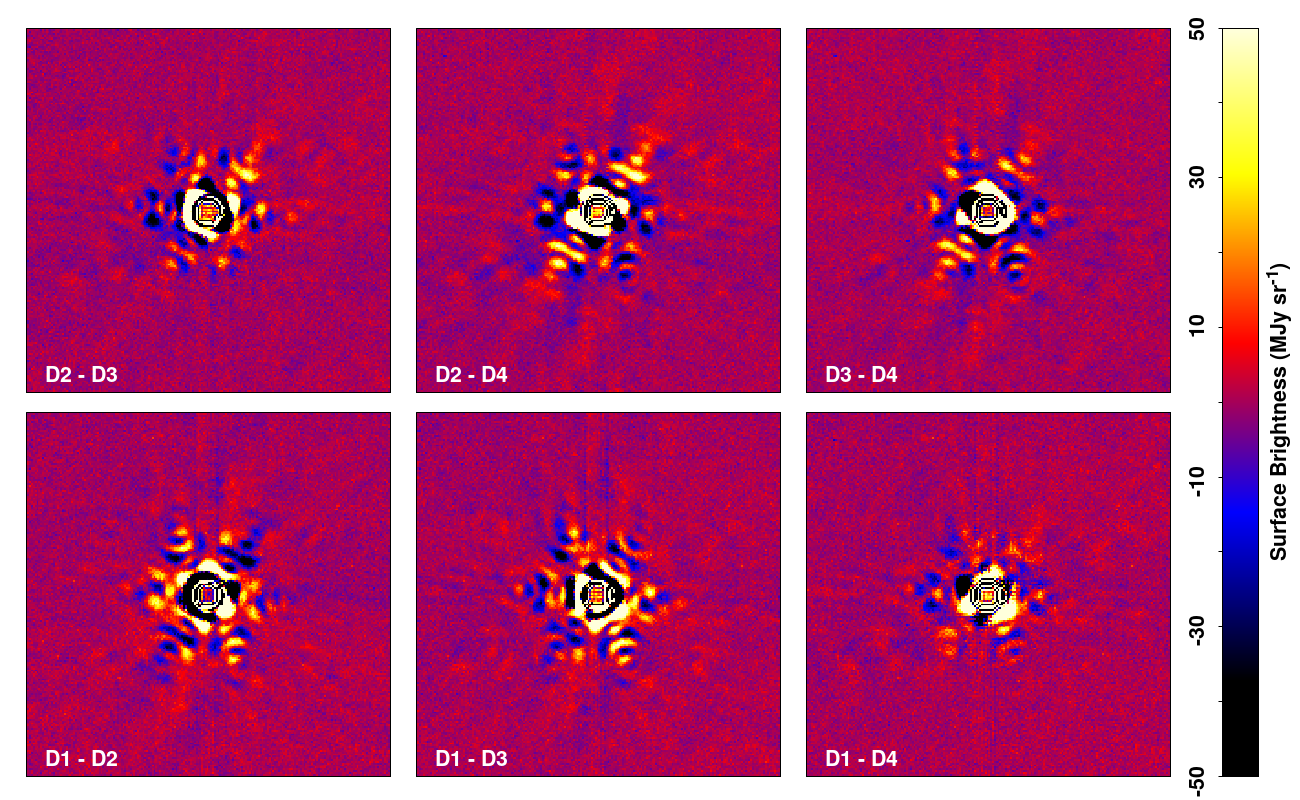}
    \caption{We show all possible
    subtraction combinations of the 4 dither positions (D1 to D4). The postage-stamp images 
    (22$^{\prime\prime}\times$22$^{\prime\prime}$) are scaled between -50 and +50 MJy sr$^{-1}$, 
    which is similar to the disk signal level at these locations. Fortunately, the PSFs taken at 
    identical positions remained remarkably stable and could be used for the
    precise removal of stellar flux (see Supplementary Figure \ref{fig:TmP}).}
    \label{fig:PmP}
\end{figure*}

\begin{figure}[!h]
    \centering
    \figuretitle{Test PSF subtractions of the Fomalhaut image at 25.5 $\micron$}
    \includegraphics[width=1.0\textwidth]{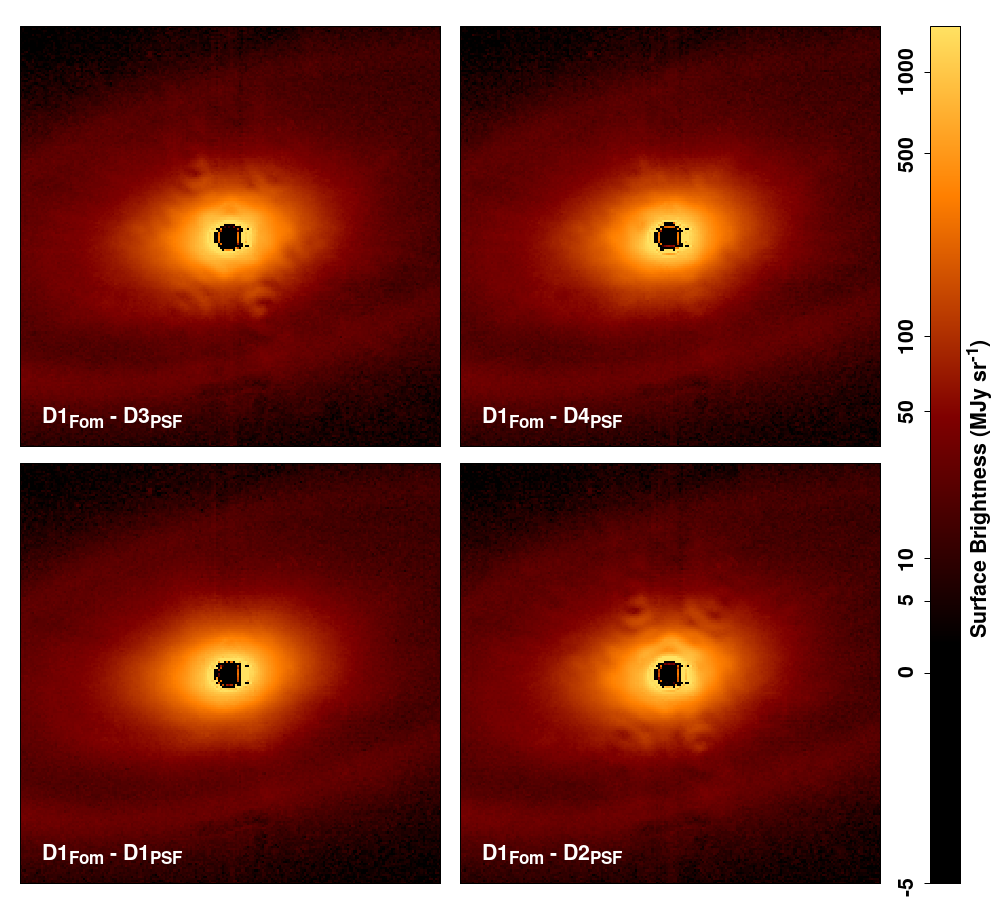}
    \caption{The test images show the PSF subtractions at the first dither position, 
    using the reference PSF observed at each of the four positions (offset from each other
    on the order of $\sim$150 px). The figure demonstrates the position dependence of the observed PSF 
    and also how well the matched position PSF is able to
    remove the stellar contribution. We ended up using only matching position PSFs for the image reductions.
    The image scaling is logarithmic between -5 and 1500 MJy sr$^{-1}$ and the image FOV is 
    22$^{\prime\prime}\times$ 22$^{\prime\prime}$. The images are oriented in the observed V3 frame of the observatory.}
    \label{fig:TmP}
\end{figure}

\begin{figure}[!h]
    \centering
    \figuretitle{The observed median radial profile of 19 PsA and the scaled theoretical WebbPSF
    model.}
    \includegraphics[width=1.0\textwidth]{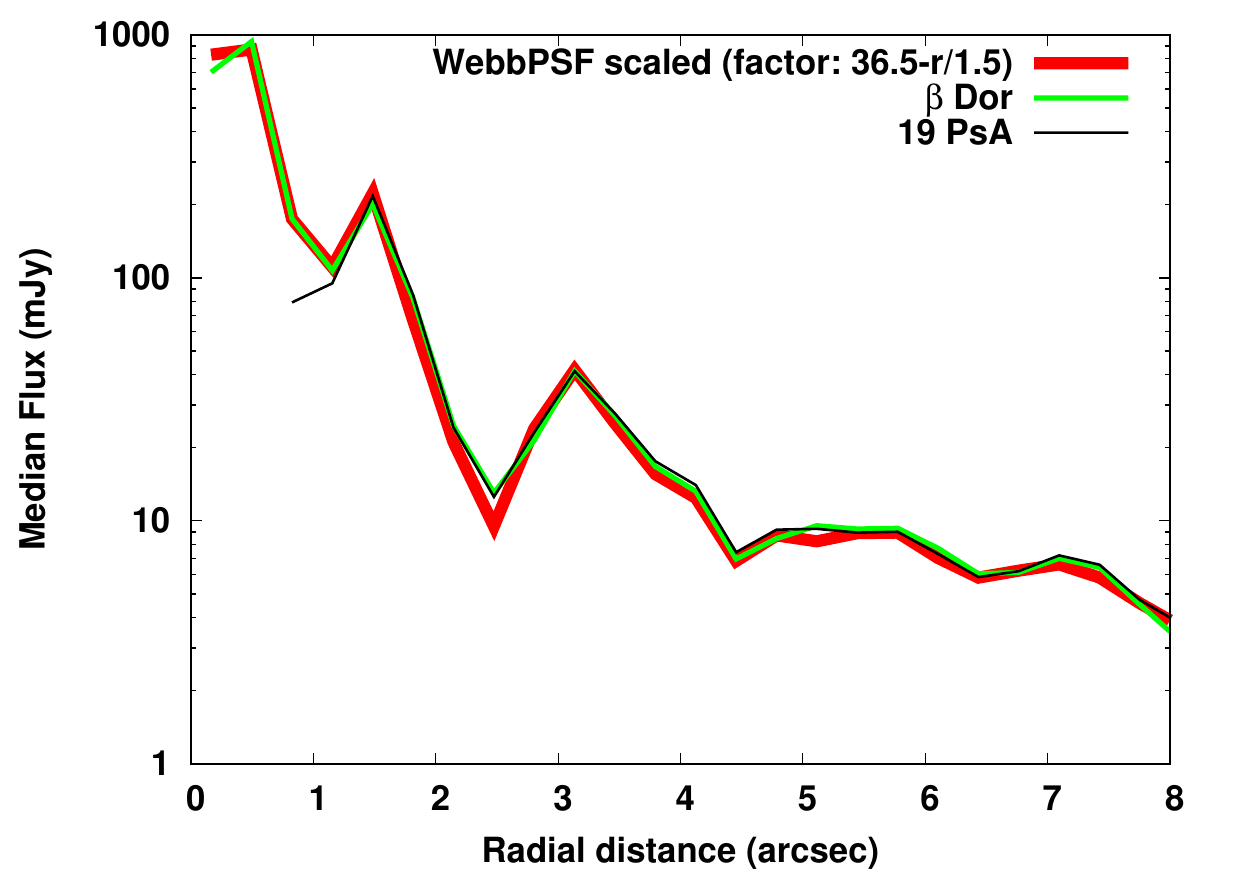}
    \caption{A radial distance dependent scaling was applied to the theoretical PSF to obtain
    a better fit at larger distances. The core of the theoretical PSF is verified with the $\beta$ Dor
    observed PSF. The total integrated flux of the theoretical PSF was only
    $\sim 10$\% fainter than the photospheric estimate for 19 PsA at 25.5 $\micron$, providing the
    scaling factor necessary for the absolute calibration of the observations. Importantly, 
    the brighter-fatter-effect that causes differences between the theoretical and observed 
    PSFs preserves charge, so the total flux should be a valid measure.}
    \label{fig:PSFfit}
\end{figure}

\clearpage
\newpage

\subsection{JWST/MIRI 23.0 $\micron$ coronagraphy}
\label{sec:lyot}

The 23.0 $\micron$ coronagraph of MIRI uses a classic Lyot mask to achieve high contrast
imaging near bright sources. The mask is rather large, providing an inner working angle 
(IWA) of $\sim 3.3\lambda/D$, i.e., $\sim 2\farcs4$ \citep{boccaletti15}. Our data is the 
first complete reference PSF subtracted set of observations taken with JWST using this mode.
\cite{boccaletti22} present raw contrasts for the F2300C coronagraph showing levels of 
$\sim 10^{-3}$ near the IWA and estimates of 10$^{-4}$ when using reference PSF subtraction
based on exposure time calculator (ETC) simulations. 

The MIRI coronagraphic fields present a scattering artifact along the Lyot mask boundaries 
-- as well as along the 4QPM mask boundaries -- known as ``glowsticks'' \citep{boccaletti22},
which are effectively removed via background subtractions. Unlike for the non-coronagraphic 
imaging at 25.5 $\micron$, dedicated background observations were taken for both Lyot and
4QPM coronagraphic imaging modes. The current status 
of the JWST pipeline, particularly the mitigation of the ``glowsticks'', results in additional noise 
because of the lack of satisfactory dark frames. Therefore, we processed the science, 
reference, and background datasets only through stage 2 of the pipeline,
producing the calibrated dataproducts (\texttt{\_calints.fits}). To use the background observations
as dark corrections simultaneously, we disabled the internal dark correction in stage 1 of the pipeline.
Our background observations for the F2300C dataset were relatively clear of artifacts. We opted to 
work with the 3D calibrated data products (which include each individual integration), instead 
of the 2D ones (which combine the integrations), as we gained a bit further flexibility in processing. 
We median combined the individual integrations within the \texttt{\_calint.fits} files for the target, 
reference, and background observations, using a 3$\sigma$ clipping algorithm centered on the data median.

Given the remarkable stability of the longer wavelength PSFs and the Lyot coronagraph's insensitivity 
to minor positional offsets, we only perform classical PSF subtraction image processing for this 
observing mode, using IRAF. We first investigated how well each individual reference PSF subtracted 
the stellar contribution from Fomalhaut by eye. The SGD of the reference 
PSF displaces the source in a 3$\times$3 grid with 10 mas (0.09 px) steps centered on the assumed 
position of the highest attenuation of the Lyot mask. The coronagraphic target acquisition 
is precise to 5 mas (0.045 px) \citep{boccaletti22,rigby22}, therefore it is highly likely that at least one of the SGD 
reference observations is close to identical to the science target. This high precision matching
is less important for the Lyot mask than it is for the 4QPM coronagraphs, where the PSF structures
are highly sensitive to positional offsets. The Lyot PSF is practically identical within the coronagraphic pointing 
accuracy of the observatory and SGD may not be necessary for observations within this mode. 
To increase the signal-to-noise ratio of the final product, we include all observed PSFs in 
our image processing steps, not just the best centered one. Determining the 
$\sim 0.05$ pixel offsets between the science and reference frames 
was complex, requiring us to track the position of the central light peak, a result of either 
light leaking through the Lyot mask or a diffraction artifact. The scaling for the reference 
PSF was set to a factor of 0.87, which resulted visually in the least amount of subtraction 
residuals. Subtraction artifacts started to become apparent at scaling factors of 0.82 and 0.92. 
Given the large occulted inner region and the high luminosity of the disk at this wavelength,
relative to the star, this large range of acceptable PSF scalings is not surprising. 

\begin{figure}
    \centering
    \figuretitle{The F$_{23}$/F$_{25.5}$ flux ratio as a function of 25.5$\micron$ brightness}
    \includegraphics[width=1.0\textwidth]{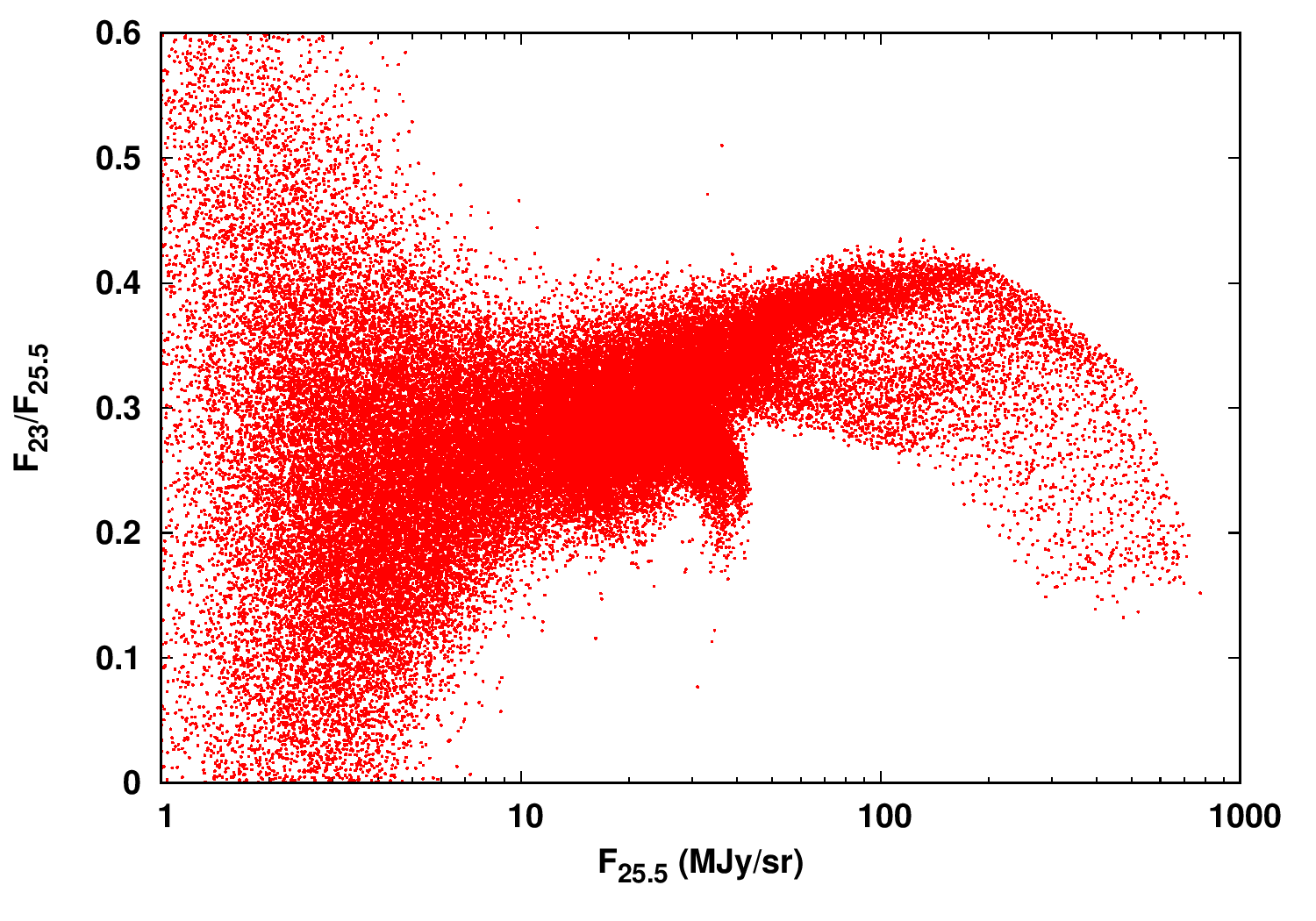}
    \caption{The F$_{23}$/F$_{25.5}$ flux ratio is shown on a per-pixel basis, in overlapping
    image regions. We would anticipate a flat line near 0.9-1, given Spitzer IRS observations of the 
    system (see Section ``The Asteroid-belt analog'') and the near identical observational wavelengths.}
    \label{fig:F25vsF23}
\end{figure}

\begin{figure*}
    \centering
    \figuretitle{The JWST/MIRI F2300C 23.0 $\micron$ Lyot coronagraphic observations of the 
    Fomalhaut debris disk system.}
    \includegraphics[width=1.0\textwidth]{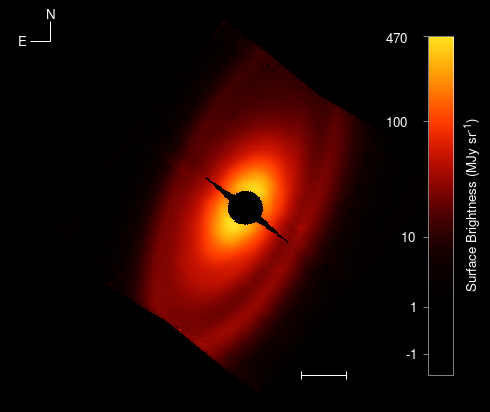}
    \caption{The image is displayed in logarithmic scaling between 
    -1.5 and 470 MJy sr$^{-1}$. The intermediate belt and the secondary gap, previously
    identified in the F2550W image, are also present in the F2300C image, as well as the
    extended inner disk.}
    \label{fig:F2300C}
\end{figure*}

Our 23.0 $\micron$ images presented a brightness dependent scaling offset in flux
values relative to the 25.5 $\micron$ observations, where the images overlapped. 
In Supplementary Figure \ref{fig:F25vsF23}, we show the F2300C/F2550W flux ratio, as a function of 
the F2550W brightness, on a per pixel basis. Given the similar wavelengths of the two observations, 
this was unexpected. The increase in the 23.0 to 25.5 $\micron$ flux ratio as a function of 
brightness could be due to issues with linearity correction, but the global scaling offset points 
to suspected issues with the pipeline (v.\ 1.8.2) calibrations at 23.0 $\micron$. This is supported
by the fact that we were able to verify the flux calibration of the 25.5 $\micron$ data to within 10\%.
For our current work, we calibrated the F2300C images with the F2550W observations. Based 
on IRS spectra (see Section ``The Asteroid-belt analog'') the inner disk should be around $\sim 6\%$ brighter at 
25.5 $\micron$ than at 23 $\micron$. In the colder parts of the disk the ratio is likely 
somewhat higher. The average F$_{23}$/F$_{25.5}=0.32$ in regions where the F2550W flux
is between 10 and 100 MJy sr$^{-1}$. To calibrate the 23.0 $\micron$ data, we multiply the observations
by a factor of 3.14. We note therefore that the absolute calibration of the F2300C observations is
uncertain.

Our final image processing result for the Fomalhaut debris disk system at 23 $\micron$ is shown 
in Supplementary Figure \ref{fig:F2300C}. The features present in the F2550W image can be identified in the 
F2300C image as well. While the IWA is larger for the F2300C image, its integration ramps 
used 102 frames (groups), thereby contributing less noise than the F2550W image from
ramp fitting alone. 

\subsubsection{JWST/MIRI 15.5 $\micron$ coronagraphy}

The 15.5 $\micron$ coronagraphic mode of MIRI uses a four-quadrant binary phase mask (0, $\pi$)
to suppress the stellar contribution to the observed signal
\citep[4QPM;][]{rouan00,boccaletti04,boccaletti15}. The on-sky performance of the coronagraph has been
excellent \citep{boccaletti22}, achieving contrasts of $\approx 10^{-4}$ within 1$^{\prime\prime}$, 
following post-processing.

Our image reduction steps for these images, up to stage 2, were identical to those used for the Lyot images; i.e.\ we
turned off dark subtractions and processed the images through the standard JWST pipeline otherwise.
We did not, however, follow the standard steps for the background subtractions for the F1550C 4QPM observations.
Our background images for both the target and reference source did not employ dithers and were contaminated with
bright PSF structures from sources outside the FOV. We obtained cleaner background data 
from the archive, utilizing observations taken within the early release science program PID 1386 
(observations 30, 36, and 37). These observations integrated to larger group numbers than our observations,
therefore we truncated them at the same levels as ours to provide similar dark noise characteristics. 
We also utilized flat-field observations taken in commissioning program PID 1045 (observation 66), using them as 
background data, for our PSF reference observation. This particular dataset yielded a background image that agreed
better with our original PSF background observation, as determined by examining residuals. Background level offsets 
were determined and corrected using median pixel values in matching clear areas. While the final reduction using the 
archival background images contained less artifacts, it is characteristically and photometrically identical to the 
version using the dedicated background observations. Following the removal of the background from the 2D PSF and 
Fomalhaut images, the ``glowsticks'' were effectively removed. 

\begin{figure*}
    \centering
    \figuretitle{The JWST/MIRI F1550C 15.5 $\micron$ 4QPM coronagraphic observations of the inner regions of the Fomalhaut ABA disk.}
    \includegraphics[width=1.0\textwidth]{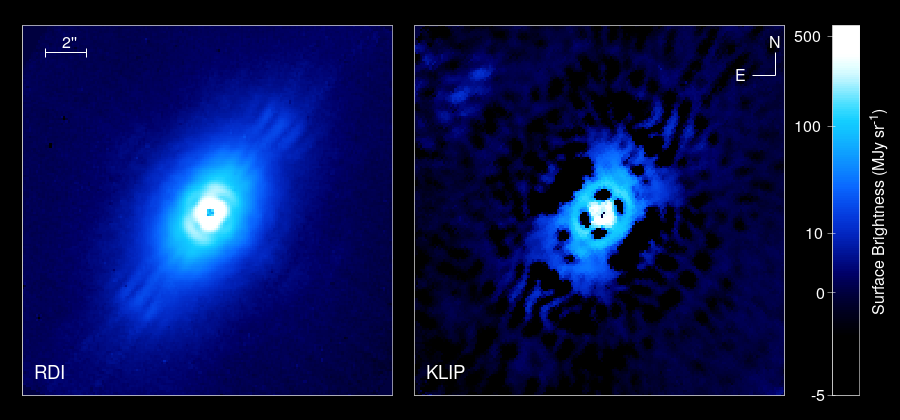}
    \caption{
    An asymmetric extended inner disk is detected in the observations, mostly aligned with the outer KBA narrow ring, 
    which is not detected. While our pre-launch models assumed a narrow inclined ring -- similar to the KBA ring -- the
    observations show that the inner ABA region of Fomalhaut is rather extended and has low surface brightness. 
    Unfortunately, our observations were oriented such that the position angle of the ABA disk aligned with the 
    PSF artifacts of the 4QPM. Image scaling is between -5 and 600 MJy sr$^{-1}$.}\label{fig:F1550C}
\end{figure*}

\cite{boccaletti22} show that classic RDI yields comparable results to the Karhunen-Loeve 
Image Processing algorithm \citep[KLIP;][]{soummer12} at 15.5 $\micron$; therefore, 
our team processed the coronagraphic observations using both classical Reference Differential Imaging (RDI)
as well as Principal Component Analysis (PCA) based on combined reference PSF subtractions employing KLIP. Below, we give a summary of
our processing steps using these two different methods. In Supplementary Figure \ref{fig:F1550C} we show the results of the
two reduction methods for the 15.5 $\micron$ 4QPM observations of the Fomalhaut system. We did not pursue 
ADI image processing, as the 10$^{\circ}$ rotational dither (the limit the observatory can rotate around its 
nominal orientation) did not provide adequate offset at such small inner working angles.

First, we describe our classical RDI image processing, all performed using IRAF. Our processing steps were
similar to those employed for the F2300C mode, with one notable difference. While we fitted PSF offsets
at F2300C, for the 4QPM PSF subtractions we purposely did not apply any positional shifts. As the 
4QPM PSF structures are position dependent, shifting the PSFs to obtain a better subtraction is mostly 
a futile exercise (either the subtraction works without shifts applied or it does not at all). For PSF 
scaling, we investigated values near the ratio of the photospheric emission of Fomalhaut vs.\ 19 PsA, 
which is $\sim$ 0.9 at 15.5 $\micron$. We noticed clear over-subtraction residuals at a factor of 0.92 
(for 19 PsA) and under-subtraction at 0.85, therefore settled on the factor of 0.88. The best 
fitting PSFs (ones with the least residuals following subtractions) were removed from the target 
observations at both orientations. For the Fomalhaut Obs \#9 image we used the PSF taken at SGD 
position \#2 and for the Fomalhaut Obs \#11 we used the PSF taken at SGD position \#1. Following 
masking and image derotations, the two images were averaged.

The JWST pipeline has a KLIP implementation included in stage 3 (\texttt{calwebb\_coron3}), which first 
performs outlier rejection and determines the reference PSF alignment via least squares. The main limitation of 
this KLIP variant is the lack of user input; the only parameter the user can define is the number of KL modes 
to use in the subtraction (the default is 50). For this reason, we also tested the spaceKLIP software developed 
by the JWST High Contrast ERS team (PI Sasha Hinkley), which is much more interactive and contains built in 
analysis tools; for more details on spaceKLIP see \cite{kammerer22} and \cite{carter22}. Ultimately, we 
found that, for extended disk observations where a single subtraction zone is optimal, spaceKLIP and 
the JWST pipeline KLIP produce equivalent results. In this work, we elected to use the JWST Pipeline KLIP 
variant for simplicity, resuming the processing of the images following our custom background subtraction
to stage 3. We performed KLIP processing with a single KL mode. For the reference library, we 
used the entire set of nine SGD images obtained for the reference target.

Our tests show that the classical RDI methods yield smoother reductions than KLIP for extended sources 
observed with the F1550C 4QPM. This is mostly due to the over-subtraction with KLIP (which 
becomes worse as more KL modes are included), as it is unable to differentiate between stellar and 
extended brightness features. In our analysis we used our classical RDI processing results.

\subsection{Cycle 27 HST Observations}

We were awarded 8 orbits of HST time in Cycle 27 (GO 15905) to observe the ABA disk around Fomalhaut in 
scattered light using STIS. Previous coronagraphic observations had a relatively large inner working angle (IWA) due to use of the 
WEDGE2.5 occulting positions. Our observations were designed to yield a  360$^{\circ}$ image of the inner regions
down to IWAs of $0\farcs3$. To accomplish this, we used two coronagraphic positions within the same visits (orbits), 
which were located on the perpendicularly intersecting  WedgeA and WedgeB occulting bars. The small IWA was achieved with the standard WedgeA0.6 position
and its unofficial pair on the WedgeB bar that is reached by using additional POSTARG commands from the WedgeB1.0 aperture. This 
observing method has been previously successfully executed by \cite{apai15} to observe the $\beta$ Pictoris debris disk (GO 12551). 
With the two Wedge positions, we achieved adequate rotational dithers and a full roll coverage of the system. Our observations were 
split into two groups of four,
due to scheduling and rotational constraints; therefore, the initial planned dataset consisted of 2 reference and 6 target observations.
Unfortunately, the PSF observations failed in the first epoch, therefore the entire first set of observations was repeated. 
In Supplementary Table \ref{tab:HSTobs}, we summarize the HST dataset.

\begin{table}
\centering
\caption{Multi-roll {\it HST} coronagraphic observations of the Fomalhaut system at WEDGEA0.6 and WEDGEB0.6 \label{tab:HSTobs}}
\begin{tabular}{lccccc}
Target             & Date             & PA orientations ($^{\circ}$)         & Time  & $N_{\rm WA0.6}$   & $N_{\rm WB0.6}$ \\
\hline\hline
Fomalhaut          & Aug-23-2020      & -83.61, -86.69, -103.36 & 27.0 s        & 45$\times$3            & 45$\times$3    \\
\hline
Fomalhaut          & Nov-18-2020      &  40.05, 24.45, 10.61    & 28.5 s        & 45$\times$3            & 50$\times$3    \\
$\Theta$ Peg (PSF) & Nov-18-2020      &                         & 27.0 s        & 45$\times$3            & 45$\times$3    \\
\hline
Fomalhaut          & Aug-20-2021      & -83.70, -86.57, -103.20 & 27.0 s        & 45$\times$3            & 45$\times$3    \\
$\Theta$ Peg (PSF) & Aug-20-2021 &                              & 27.0 s        & 45$\times$3            & 45$\times$3    \\
\end{tabular}
\end{table}

\begin{figure*}
    \centering
    \figuretitle{The inner regions of the Fomalhaut system with the HST Cycle 27 (GO 15905) observations. }
    \includegraphics[width=1.0\textwidth]{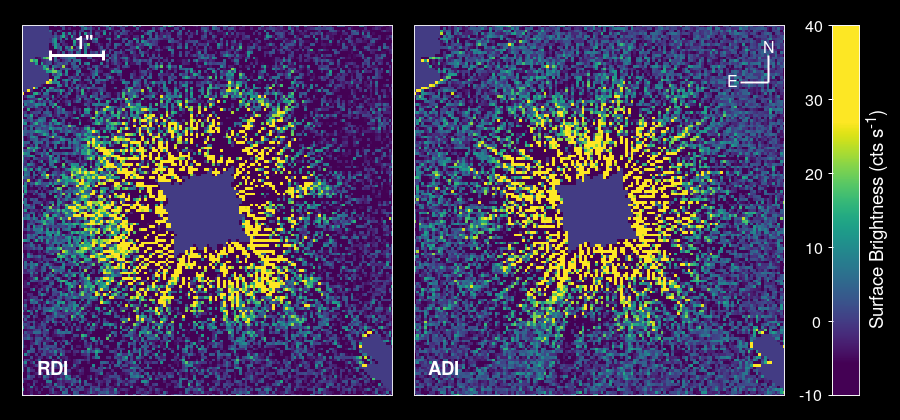}
    \caption{The left panel shows the RDI reduction while
    the right panel shows the ADI reduction results, using classical PSF subtraction methods. Both images are displayed in linear scaling between
    -10 and 40 cts s$^{-1}$. The observations/reductions are contrast limited out to 3$^{\prime\prime}$ and we do not detect any discernible levels
    of extended scattered light emission at any location.}
    \label{fig:HSTobs}
\end{figure*}

\begin{figure}
    \centering
    \figuretitle{The Median Absolute Deviation contrast achieved with the HST Cycle 27 (GO 15905) observations. }
    \includegraphics[width=1.0\textwidth]{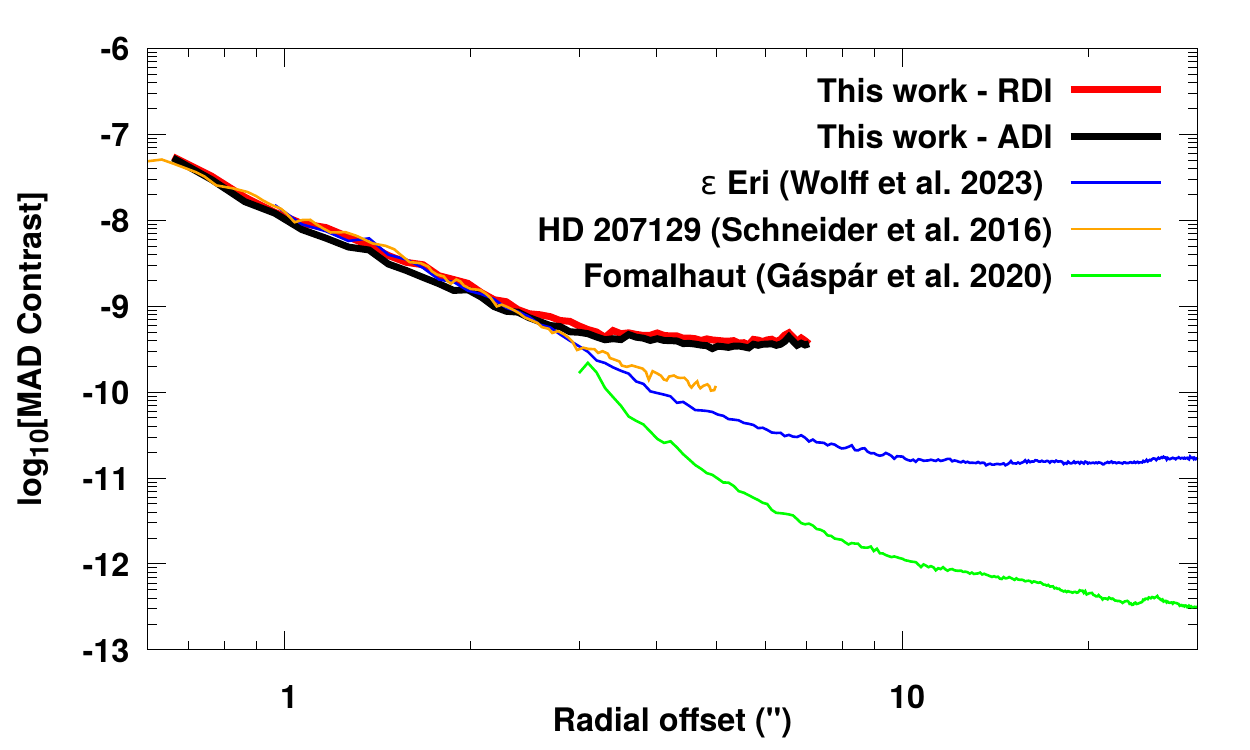}
    \caption{Given the short total integration time of 
    less than 1.5 minutes - a result of the planned small IWA of the observations - our achieved contrast levels taper off to being photon limited
    at around 3$^{\prime\prime}$.}
    \label{fig:HSTcontr}
\end{figure}

Reductions of the datasets followed along the lines of our $\epsilon$ Eridani observations \citep{wolff23}, where we first removed the video-noise
inherent in the STIS data using the autofillet idl package \citep{jansen03} and then reduced the corrected raw data with the calstis pipeline.
These HST observations were amongst the most challenging, in terms of image processing, ever taken. Since Fomalhaut was near saturation at the
WA0.6 aperture, we recorded 820 individual frames of Fomalhaut alone, and 175 of its reference calibrator. To lessen the workload, we first median 
combined the individual frames in groups of five (in sequential order), 
thereby having only 164 Fomalhaut frames to individually cross-reference with 35 PSFs. Numerical algorithms are not as accurate as the human eye
in finding optimal alignment and scaling between target and reference data; therefore, we first found the best fitting PSF for each
observed target observation using an initial best-guess of its center location with centerRadon \citep{ren19}. Once the best matching PSF was found
for each target observation, they were individually re-aligned. While on average centerRadon works well, we found small adjustments necessary
to achieve ideal subtractions. At WA0.6 our average shifts were $\Delta x = -0.0177 \pm 0.0321$ and $\Delta y = 0.00556 \pm 0.0304$ pixels, while at
WB0.6 they were $\Delta x = 0.00569 \pm 0.0566$ and $\Delta y = -0.0389 \pm 0.0787$ pixels. While these may seem negligible, at 0.05 px offset subtraction
speckle residuals are the major source of noise; therefore, this was a necessary step. Finally, we also adjusted the scaling of the PSFs for each matched
and aligned combination to produce the same signal level. Following masking, image translations/rotations, and finally combining the images via median averaging
(and iterative sigma clipping around the data median), our image processing work resulted in the null detections we show in Supplementary Figure \ref{fig:HSTobs}.
In the figure, we also present a reduction using ADI methods, where we performed the same operations as above, just cross-refencing the Fomalhaut observations
with themselves, avoiding the orbits where the orientation angles were within $10^{\circ}$ of each other. While the observations totalled only less
than 1.5 minutes of integration time on target, the contrast levels we achieved are nonetheless impressive, especially compared to other, much
longer integration programs, as shown in Supplementary Figure \ref{fig:HSTcontr}. Our combined dataset is contrast limited to around 3$^{\prime\prime}$.
Within that region, however, our reductions are on-par with programs using significantly longer integrations, due to the absolute brightness
of Fomalhaut and the remarkable stability of the HST optical system. However, these observations cannot compete with the contrast levels 
achieved in \cite{gaspar20}, which combines over 8 hours of STIS data using much larger coronagraphic apertures (thereby imaging only the 
outer regions of the system).

The lack of a scattered light detection with HST may seem surprising, given the bright warm component observed in the MIRI data. Our assumption
that the component could be observed with HST was based on the spatially unresolved warm excess with Spitzer \citep{stapelfeldt04,su13}. 
These data hinted at a compact inner disk, one very similar to the Solar System asteroid belt. The JWST images show that the warm component is 
spatially extended, thereby having a much lower scattered light surface brightness than we expected. The JWST and HST observations together 
enable complex modeling of the dust grain properties and light scattering functions, which we will address in future papers.

\section{Analysis: Supplementary Information}

\subsection{Photometry}

Photometric analysis was performed on the images in two configurations: in the observed
frames and also on de-projected images. Photometry in the observed frames allow for comparisons with previous measurements, 
where the inner regions were not resolved spatially, while photometry of the de-projected 
images yields the input for modeling the system. Statistical errors were calculated using the error frames of the observations,
while systematic errors from PSF fitting were estimated using the scaling ranges we
determined with by-eye fitting. At 25.5 and 23.0 $\micron$ we also include an additional
conservative 5 and 10\% systematic error term in quadrature, respectively, to allow for
offsets from the absolute calibrations. The allowed error at 23.0 $\micron$ was set to be larger, 
due to the added uncertainty in the scaling factor from the 25.5 $\micron$ observation. 
The systematic errors were the dominant error terms at all wavelengths. We assume the
systematic error terms of these observations to improve over time, as the behavior of 
the MIRI imager at these wavelengths is better characterized.

\subsection{Deprojection}
\label{sec:deproject}

\begin{figure}
   \centering
   \figuretitle{Flux contour lines at 15.5 $\micron$ (blue) and 25.5 $\micron$ (red). }
    \includegraphics[width=0.45\textwidth]{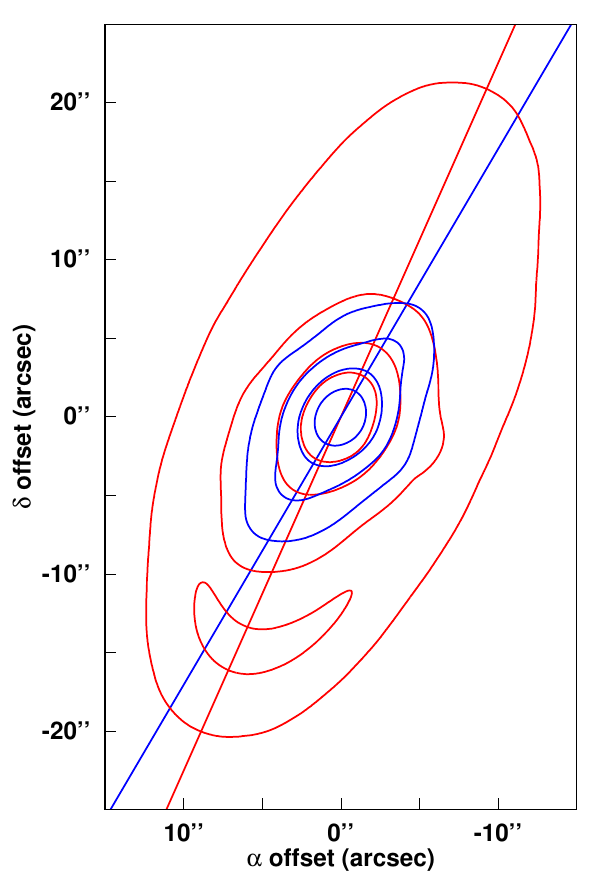}
    \caption{The F1550C lines are drawn at
    3.16, 10, 31.6, 100, and 316 MJy sr$^{-1}$, while the F2550W lines are drawn at 10, 31.6, 100, and 316 MJy sr$^{-1}$.
    Vector lines are drawn following along the position angle of the disks (336$\fdg$04 and 329$\fdg$59), to highlight 
    the offsets between the inner and outer regions. Note that the F1550C and the F2550W contours align well within the ABA region.}
    \label{fig:contours}
\end{figure}

\addtolength{\tabcolsep}{0pt}
\begin{table}[]
\centering
\caption{The fitted orbital parameters of the KBA inner gap and that of the contour levels shown in Supplementary Figure \ref{fig:contours}.\label{tab:PAs}}
\begin{tabular}{lccccc}
Flux              &  a             &  e               &   PA              & $\iota$         & $\omega$     \\
(MJy sr$^{-1}$)   & (au)           &                  &  (deg)            & (deg)           & (deg)        \\
\hline
\multicolumn{6}{c}{KBA inner gap - used for de-projection}\\
\hline\hline
KBA gap           & 133.79$\pm$0.26 & 0.132$\pm$0.002 &  336.28$\pm$0.09  &  67.52$\pm$0.09 & 38.04$\pm$0.8 \\
\hline
\multicolumn{6}{c}{F1550C}\\
\hline\hline
3.16              & 66.38$\pm$0.08 & 0.037$\pm$0.001 & 326.22$\pm$0.09  &  56.94$\pm$0.08 & 170.72$\pm$1.75   \\
10.0              & 44.63$\pm$0.09 & 0.028$\pm$0.002 & 326.50$\pm$0.17  &  56.19$\pm$0.15 & 161.42$\pm$4.10   \\
31.6              & 25.86$\pm$0.11 & 0.022$\pm$0.004 & 327.79$\pm$0.62  &  44.17$\pm$0.43 & 142.56$\pm$9.80   \\
100.0             & 14.63$\pm$0.14 & 0.020$\pm$0.008 & 328.72$\pm$2.01  &  36.23$\pm$1.23 & 161.51$\pm$24.03  \\
\hline
\multicolumn{6}{c}{F2550W}\\
\hline\hline
10.0              & 172.48$\pm$0.05 & 0.020$\pm$0.001 & 336.04$\pm$0.02  &  64.42$\pm$0.02 & 5.42$\pm$0.85   \\
31.6              & 74.35$\pm$0.07  & 0.131$\pm$0.001 & 326.71$\pm$0.09  &  52.07$\pm$0.08 & 176.37$\pm$0.43 \\
100               & 40.23$\pm$0.09  & 0.045$\pm$0.002 & 329.59$\pm$0.28  &  47.80$\pm$0.21 & 149.65$\pm$2.61 \\
316               & 23.36$\pm$0.12  & 0.045$\pm$0.004 & 332.01$\pm$0.77  &  43.09$\pm$0.51 & 129.52$\pm$5.50 \\\\
\end{tabular}
\begin{flushleft}
\footnotetext{
Semi-major axis lengths were calculated using a pixel scale of 0.11$^{\prime\prime}$/px and a distance of 7.7 pc. The fits and joint errors
were determined with {\texttt emcee} using the Markov chain Monte Carlo fitting algorithm assuming Keplerian orbits. The KBA gap fit results
were used to de-project the images.}
\end{flushleft}
\end{table}

In Supplementary Figure \ref{fig:contours}, the flux contour profiles of the disks at 
15.5 and 25.5 $\micron$ are compared, chosen at logarithmic levels. We forego the contour analysis of
the 23.0 $\micron$ data, as it is clipped both at the outer and inner regions and it is characteristically 
the same as the 25.5 $\micron$ data where they overlap. Keplerian orbits were fitted to these contours 
via Markov chain Monte Carlo (MCMC) fitting methods, using {\texttt emcee}, allowing for
joint Bayesian errors to be estimated. The orbital parameters were determined using 6000 test
chains with 10000 steps, allowing for a burn-in limit of 200 steps. Additionally, we determined
the orientation of the system by fitting 51 points chosen by eye at the inner edge of the KBA ring 
with the same orbit fitting method as for the contours. This was necessary, as the contour levels 
track locations of equal illumination, which do not necessarily correspond to orbital paths due to the 
higher eccentricity and inclination of the KBA ring. The system orientation determined using 
the outer ring inner gap is similar to that calculated based on ALMA data \citep{macgregor17},
although with an inclination higher by $\sim 2^{\circ}$. We summarize the inner gap and contour fits in Supplementary Table \ref{tab:PAs}.

The fitting reveals an inner disk mostly aligned with the outer KBA ring; however, there is a gradual
misalignment between the disk components, with the inner disk offset by $\sim 6.7^{\circ}$
in position angle relative to the KBA ring. The outer ring has a position angle of 
$\sim 336^{\circ}$ and inclination of 64.4$^{\circ}$ (determined at the 10.0 MJy sr$^{-1}$
contour), while the inner disk is oriented at $\sim 329.6^{\circ}$ with a much smaller inclination angle 
of $47\fdg8$. In general, the position angles of the inner components are smaller than for the outer with
the inclination angles also decreasing inwards. The inner disk alignment agrees at 15.5 and 25.5 $\micron$, 
demonstrating that the misalignment is not due to the PSF residuals of the 4QPM. The outermost 15.5 $\micron$ 
contour profile is affected by the 4QPM PSF residuals, therefore any further misalignment it presents was ignored.
Gravitational perturbations by a massive planet within the disk system could have plausibly carved the secondary disk gap in the 
system and forced this misalignment. 

\subsection{The Kuiper-belt analog}
\label{sec:KBA}

The most extended component of the Fomalhaut debris disk system is its large and narrow KBA ring. 
First resolved in scattered light with HST/ACS \citep{kalas05}, the outer ring has since been imaged at many 
wavelengths. Here, we report only on the observations of this component with JWST/MIRI at 23 and 25.5 $\micron$ as it
was not detected in the observations at 15.5 $\micron$.
As measured on the deprojected F2550W image, the KBA ring has a sharp inner edge with a semi-major axis of 133.8 au
and a width of approximately 20-25 au. We measure an eccentricity of $e=0.132\pm0.002$ for the inner edge, in agreement
with previous estimates \citep[e.g. $e=0.12\pm0.01$;][]{macgregor17}. Given Fomalhaut's 16.63 L$_{\odot}$ 
luminosity, the thermal location of the KBA ring corresponds to an equivalent of 30 - 40 au around the Sun, 
establishing this ring as a true analog of the Kuiper belt.

\begin{figure}[!t]
    \centering
    \figuretitle{The median surface brightness of the KBA ring}
    \includegraphics[width=0.8\textwidth]{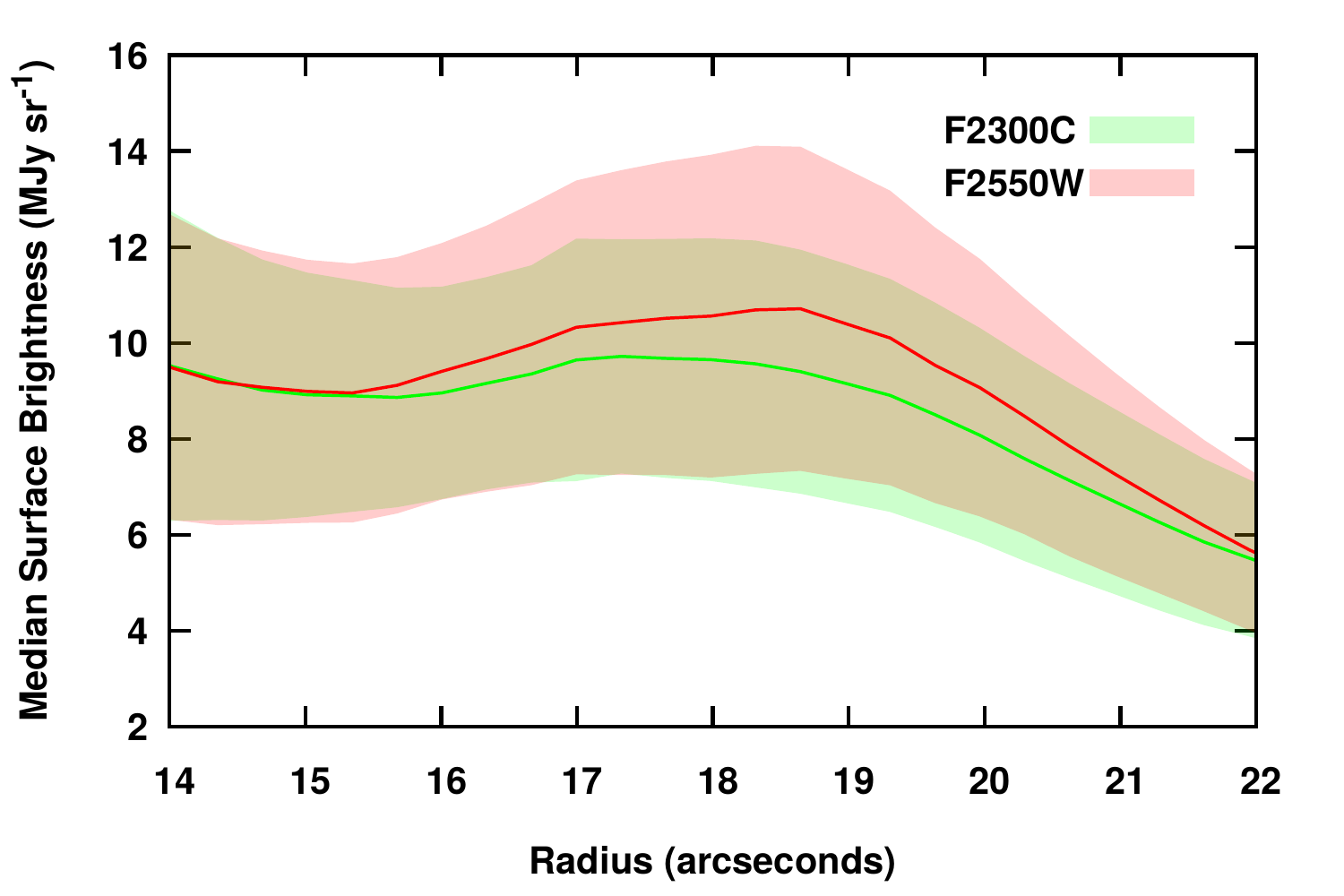}
    \caption{The median surface brightness is measured on the deprojected images, centered on coordinates 
    22$^{\rm h}$57$^{\rm m}$39$\fs$62, -29$^{\circ}$37$^{\prime}$22$\farcs$06 (not on Fomalhaut) 
    at the time of the observations. Statistical errors were negligible. The dominant systematic errors are 
    based on PSF subtraction scalings with a conservative 5\% and 10\% additional error added in quadrature at 
    25.5 and 23.0 $\micron$, respectively.}
    \label{fig:axis}
\end{figure}

We show the median radial profile of the ring in Supplementary Figure \ref{fig:axis} at 23.0 and 25.5 $\micron$, using the de-projected
image shown in main text Figure 2. We determine the radial profile centered on coordinates 
22$^{\rm h}$57$^{\rm m}$39$\fs$62, -29$^{\circ}$37$^{\prime}$22$\farcs$06,
which is the center of the KBA ring in the deprojected image at the observation epoch (not the location of Fomalhaut). 
The median peak surface brightness of the ring is 9.7$\pm$2.4 MJy sr$^{-1}$ at 23.0 and 10.7$\pm$3.4 MJy sr$^{-1}$ at 
25.5 $\micron$. The complete disk was not imaged in the Lyot coronagraphic subarray, but we can extrapolate an estimated 
total disk brightness of 114$\pm$12 mJy at 23.0 $\micron$ and measured a value of 126$\pm$9 mJy at 25.5 $\micron$ 
between 15$\farcs$25 (117 au) and 20$\farcs$1 (155 au) from this central location in the deprojected image. These
flux values provide the specific total flux from the disk component itself to allow for detailed modeling.
An extended blow-out halo is quite apparent in both the HST scattered light and 25.5 $\micron$ 
thermal images at the two apices. In the halo, outside of $20\farcs1$, we measure a total brightness of 126$\pm$13 mJy
at 25.5 $\micron$, which is just as much as within the KBA ring itself. The majority of this halo flux originates from the regions
near the KBA ring.

\subsection{The panchromatic imaging gallery of the Fomalhaut debris disk system}
\label{sec:gallery}

\begin{figure*}[ht]
    \centering
    \figuretitle{The panchromatic gallery of Fomalhaut observations}
    \includegraphics[width=0.98\textwidth]{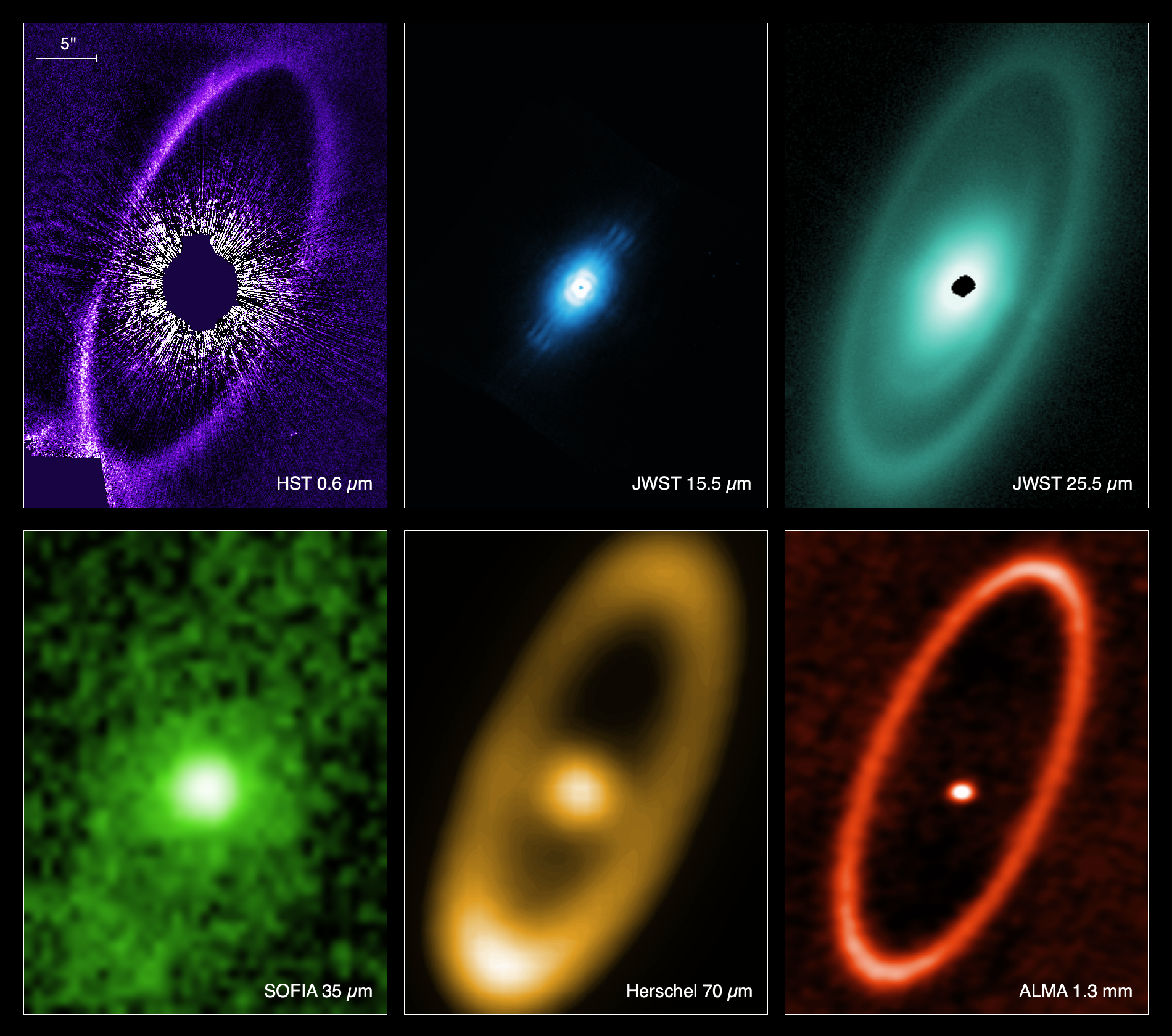}
    \caption{The images are from the following studies: HST/STIS - \citep{kalas13,gaspar20}, 
    JWST/MIRI - this work, SOFIA/FORECAST - \citep{adams18}, Herschel/PACS - \citep{acke12}, ALMA - \citep{macgregor17}. 
    The gallery shows how different wavelengths highlight different regions within the disk and how JWST is uniquely able 
    to resolve structures within the inner regions. Images are displayed at their proper orientations, with N to the top and
    E to the left.}
    \label{fig:gallery}
\end{figure*}

The latest JWST observations of the Fomalhaut system complete the panchromatic spatially resolved imaging of the system.
We would be remiss if we didn't use this opportunity to present the reader with said gallery, allowing a detailed view
of dust scattered light and thermal emission from optical, through mid-IR to radio wavelengths. In Supplementary Figure \ref{fig:gallery}
we present the images from HST/STIS \citep[][this work]{kalas13,gaspar20}, MIRI (this work), SOFIA/FORECAST \citep{adams18}, 
Herschel PACS \citep{acke12}, and ALMA \citep{macgregor17}. We note that these are not the only images of the system 
\citep[e.g.\ there are also HST/ACS images as well as Spitzer/MIPS images;][]{kalas05,stapelfeldt04}, but are the 
deepest, highest spatial resolution, and most complete observations at the given wavelengths. These images highlight 
the unique diagnostic capability of each instrument/observatory and the unparalleled opportunity we have with JWST to 
study such systems at mid-IR wavelengths at high spatial resolution.

\subsection{Re-calibration of the IRS spectrum of the inner debris system}
The IRS spectrum of the debris system in \cite{su13} shows $\sim$ 320 mJy of emission at 15 $\micron$ from the 
inner debris system, whereas the coronagraphic results at 15.5 $\mu$m reveal ``only'' $78\pm35$ mJy. As shown in 
main text Figure 5, when the spectrum is averaged over the filter bandpass, there is a very significant discrepancy. 

To explore this 
offset, we have re-evaluated the spectrum. To do so, we use photometry of the inner system. Three of 
the measurements we use were obtained with beams of diameter 6 $-$ 10$''$ \cite{bliek96,ishihara10, su16,adams18}, 
i.e., all refer to the inner system. A fourth measurement utilized the synthetic photometry on the Spitzer IRS 
low-resolution spectrum from PID 1074 (reduction from the NASA/IPAC Infrared Science Archive, IRSA) to generate 
a value at 22 $\mu$m, again with a beam of $\sim$ 8.5$''$. A summary of the results is plotted in main text
Figure 5. 

We now describe how the values were obtained. Measurements at 2 to 8 $\mu$m are reported by \cite{su13}. They 
conclude that there is no evidence for an excess between 2 and 8 $\mu$m, other than the hot excess of 
$0.88 \pm 0.12\%$ at 2 $\mu$m detected interferometrically \citep{absil09}. Photometry in the 10 $\mu$m atmospheric 
window was obtained by \cite{bliek96}. We focus on the intermediate width ($\Delta \lambda/\lambda$ $\sim$ 0.1) 
bands at 8.361, 9.787, and 12.819 $\mu$m; although measurements in the full N-band are also reported, various types 
of systematic errors should be lower in the narrower bands, particularly given the peculiar transmission profile of 
the N-band filter used. Measurements in the intermediate bands were obtained on six nights, and the averages have 
indicated errors $\le$ 1\%. The magnitudes quoted for these measurements are based on solar colors, calculated 
from the solar spectrum in \cite{labs70}. These colors are rather far from current understanding, so we computed 
colors across the three intermediate bands using the spectra of Sirius and the Sun in \cite{rieke23}. Correcting 
to these colors and adopting the given magnitude at the shortest band, the magnitudes are: 0.916 (8.361 $\micron$), 
0.890 (9.787$\micron$), and 0.903 (12.819 $\micron$). We average the first two to get 0.901 and assume, based on the IRAC 
measurements, that there is no excess in those bands. The value in the third band then shows no excess there either. 
Given the nominal error of 3\% but the relatively large number of measurements in the intermediate bands and their 
resistance to systematic errors, we assign an error of 2\%. 

Our measurement of the Fomalhaut debris disk contribution at 18 $\mu$m is based on a self-calibrating use of the 
9 and 18 $\micron$ photometry from Akari. We first used \cite{su06} to identify A-stars with no evidence for excess 
emission at 24 $\micron$ and that are bright enough to be measured at high signal-to-noise in the two Akari bands. 
The stars passing these two tests  are HD 11636, 76644, 80007, 87901, 103287, 108767, 
112185, 130841, 135742, 209952, and 216627. We assumed all of these stars had the identical ratio of 9 to 18 $\mu$m flux densities. 
In addition to the tabulated statistical errors, we assigned an error parameter for additional errors in the 
photometry. We found that, if this additional error was set to 2.5\%, that $\chi^2$ for the average was 1. We 
took the weighted average of the determinations with this additional error. When compared with the average of 
the fluxes in the two bands for Fomalhaut, this implies an excess for it at 
18 $\micron$ of $280 \pm 95$ mJy.

We used a technique similar to that for the Akari data to derive the 22 $\mu$m result. The extraction aperture 
for the IRS spectrum (from PID 1074, IRSA) and hence the synthetic photometry based on it is $\sim$ 8.5$''$, 
closely comparable to that for the measurements at 12.8 and 24 $\mu$m.  We determined the average ratio of the 
22 $\mu$m to 16 $\mu$m photometry for A-stars without excess \citep{sloan15}, namely HR 1251, 3799, 4138, 4801, 
6789, and 7950. Making a small correction for excess in the 16 $\mu$m band, the excess above the stellar 
photospheric output at 22 $\mu$m compared with these non-excess stars is indicated to be $13.8 \pm 2.8$\%, 
or a flux of $540 \pm 110$ mJy. 

Measurements at 24 and 37 $\micron$ were taken from \cite{su16} and \cite{adams18}. Except for the last case, all
were adjusted to the calibration of \cite{rieke23}. We then used these four measurements to renormalize the 
spectrum. In addition to the statistical errors for the spectral points, we assumed an error of 0.5\% of the stellar 
photosphere at 10 $\mu$m, increasing linearly with wavelength (as $\lambda$/10 $\mu$m) toward longer wavelengths, 
to allow for uncertianties in the stellar SED \citep[e.g., due to the hot dust emission,][]{absil09,mennesson13}. 
The normalization is achieved by adjusting the calibration of the spectrum before subtracting the stellar 
continuum. Because one is subtracting a large number (the spectrum of star plus excess) from another large one 
(the photospheric emission) to get a smaller one (the spectrum of the excess), the normalization is very sensitive 
to the overall calibration of the spectrum. The result needs to be consistent with the four shorter-wavelength 
photometric points and to not go too far negative at the shortest wavelengths near 10 $\micron$.

Nonetheless, the derived excess spectrum does go somewhat negative short of 13 $\mu$m. As an independent test,
we have used the IRS spectrum from PID 1074; although it is saturated at wavelengths less than 10 $\mu$m, it is 
valid at the longer wavelengths. We normalized this spectrum so that, after subtracting the photospheric 
continuum, the average excess indicated between 10 and 14 $\mu$m is equal to zero. As shown in main text
Figure 5, the derived excess spectrum shows no gradient over this wavelength range, 
supporting the assignment of no excess (since any plausible excess would have a redder spectrum than the 
stellar photospheric one). Furthermore, the spectrum of the excess agrees closely with that from the 
renormalized one from \cite{su13}. (Beyond 30 $\micron$, the PID 1074 spectrum becomes more noisy and as 
indicated in the figure departs systematically from the spectrum from \cite{su13}, which overall 
should be the more reliable of the two.)

\clearpage
\newpage


\end{document}